\documentclass[aps,pre,twocolumn,showpacs]{revtex4}

\usepackage{amsfonts}
\usepackage{epic,eepic}
\usepackage{graphicx}

\renewcommand{\phi}{\varphi}
\renewcommand{\rho}{\varrho}
\renewcommand{\epsilon}{\varepsilon}
\renewcommand{\theta}{\vartheta}

\begin{document}

\title{Quantum Multibaker Maps: Extreme Quantum Regime}

\author{Daniel K. W\'{o}jcik}
\email{danek@ipst.umd.edu}
\homepage{http://www.cft.edu.pl/~danek}
\affiliation{Institute for Physical Science and Technology, \\
  University of Maryland, College Park, MD, 20742, USA}
\affiliation{Centrum Fizyki Teoretycznej Polskiej Akademii
  Nauk,\\ Al. Lotnik\'{o}w 32/46, 02-668 Warszawa, Poland; \\ }

\author{J. R. Dorfman}
\affiliation{Institute for Physical Science and Technology,\\ and Department
of Physics \\
University of Maryland, College Park, MD, 20742, USA}

\pacs{05.60.Gg, 05.60.-k, 03.65.-w}

\date{\today}

\begin{abstract}
  We introduce a family of models for quantum mechanical,
  one-dimensional random walks, called quantum multibaker maps (QMB).
  These are Weyl quantizations of the classical multibaker models
  previously considered by Gaspard, Tasaki and others. Depending on
  the properties of the phases parametrizing the quantization, we
  consider only two classes of the QMB maps: uniform and random.
  Uniform QMB maps are characterized by phases which are the same in
  every unit cell of the multibaker chain.  Random QMB maps have
  phases that vary randomly from unit cell to unit cell.  The
  eigenstates in the former case are extended while in the latter they
  are localized.  In the uniform case and for large $\hbar$, analytic
  solutions can be obtained for the time dependent quantum states for
  periodic chains and for open chains with absorbing boundary
  conditions. Steady state solutions and the properties of the
  relaxation to a steady state for a uniform QMB chain in contact with
  ``particle'' reservoirs can also be described analytically. The
  analytical results are consistent with, and confirmed by, results
  obtained from numerical methods.  We report here results for the
  deep quantum regime (large $\hbar$) of the uniform QMB, as well as
  some results for the random QMB. We leave the moderate and small
  $\hbar$ results as well as further consideration of the other
  versions of the QMB for further publications.
\end{abstract}

\maketitle

\section{Introduction}
\label{sec:intro}

The quantum mechanics of classically chaotic systems, often called
quantum chaos, is by now a highly developed subject with an enormous
literature, including monographs by Gutzwiller~\cite{Gutzwiller90da},
St\"ockmann~\cite{Stoeckmann99da}, and Haake~\cite{Haake01da}, among
others.  The subject has been greatly advanced, as is usual, by
detailed analyses of simple model systems such as kicked rotors,
quantum flows on surfaces of constant negative curvature, Harper
models, and so on. Some of the central problems that have been studied
using these models include those of: (1) finding explanations for the
efficacy of random matrix theories, (2) understanding the differences
between quantum and classical transport, especially when Anderson
localization plays a role in the quantum system, (3) studying the
properties of quantum systems in the semi-classical limit, and (4)
determining the role of decoherence in producing classically chaotic
behavior of a quantum system as Planck's constant tends to zero.

\begin{figure}[htbp]
  \centering
  \includegraphics[scale=0.3]{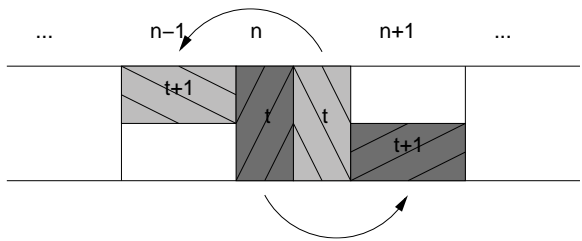}
  \caption{Classical multibaker model.}
  \label{fig:multi1}
\end{figure}

The present paper treats the quantum versions of simple model systems,
multibaker maps, that have been used to study transport phenomena in
classically chaotic systems.  A~multibaker map consists of a chain of
two-dimensional baker maps which are interconnected by means of a
simple change in the baker dynamics.  In the usual baker map on a unit
square or torus, two vertical strips are stretched (by a factor of
two) in the horizontal direction, contracted (by a factor of two) in
the vertical direction, and the resulting horizontal strips are placed
one above the other, in order to reconstruct the unit square. In the
multibaker chain, each of the two horizontal rectangles are sent to
adjacent cells, one to the right and the other to the
left~(Fig.~\ref{fig:multi1}). Modified multibaker chains have also
been studied where there may be more strips and/or a more complicated
dynamics including both area preserving and area non-preserving
dynamics. These classical models provide simple, deterministic models
of one dimensional random-walk processes with both diffusive transport
and chaotic dynamics. They have been used to study connections between
transport properties such as transport coefficients, and irreversible
entropy production, and the chaotic
properties of the models~\cite{Gaspard98sa,TasakiG00,TasakiG99,%
gilbert00s,tasaki95s,gaspard92s,gilbert99s,VollmerTM00,%
MatyasTV00,RondoniTV00,VollmerTB98,BreymannTV98,%
VollmerTB97,BreymannTV96,tel00}.

Quantum versions of baker maps are well known and studied in some
detail for a range of values of Planck's constant. Here we add a
mechanism for transport of probability amplitudes along a one
dimensional chain of quantum baker maps. This quantum version of the
multibaker map (QMB) provides one realization of a quantum random walk
process.  In this paper we will concentrate on the ``most quantum''
version of the QMB, obtained by using the largest possible value of
Planck's constant $h =1/2$, in the Weyl quantization of the ordinary
baker map~\cite{BalazsV89,Saraceno90}.  Our goal here will be to
explore the transport properties of the QMB for this value of $h$, and
for two different versions of the model, obtained by taking advantage
of some phase-related arbitrariness in the quantization of the map.
Later papers will explore further properties of QMB's including the
semi-classical case, not considered here~\cite{wojcik02de}.

Area-preserving maps on torus admit a two-parameter family of Weyl
quantizations~\cite{Hannay80,Saraceno90,DeBievreEG96}, where the two
parameters can be chosen to be phases. One can think of the two phases
as offsets of lattice points that define the spatial and momentum
coordinates of the map.  If we choose the same phases in each unit
cell of the chain, we have a ``uniform'' QMB. If we choose random
phases from cell to cell, we obtain a ``random'' QMB. The different
versions have quite different properties, as one might expect. The
uniform QMB has many features in common with those of continuous
one-dimensional systems with periodic potential, including extended
eigenstates, and ballistic transport, while the random case exhibits
the usual phenomena associated with localization. Nevertheless, there
are some interesting surprises, as we shall see in further sections,
associated with transport in open systems.

There are a number of formulations of quantum random walks already in
the literature. We mention, in particular, work of Aharonov {\it et
  al.}~\cite{aharonov93s}, work of Godoy {\it et
  al.}~\cite{GodoyF92,GodoyG96}, and work of Barra and Gaspard
\cite{barra99da}.  The papers of Godoy and co-authors as well as that
of Barra and Gaspard have interesting parallels with ours. These
authors consider the motion of a quantum particle along a
one-dimensional, periodic chain of scattering sites. The scattering
sites are characterized by transmission and reflection amplitudes,
which for a periodic system, are taken to be the same for each site.
Godoy and co-workers consider the wave functions for their systems at
discrete positions and discrete times, and propose a set of equations
similar to the ones considered here. These equations are then solved
using stationary phase approximations, and the connections with
Landauer's formula are discussed, for various parameters and particle
statistics. Barra and Gaspard also consider a model similar to ours,
and they analyze the scattering resonances for a finite, open system.
By applying transfer and $S$ matrices, they obtain expressions for the
widths of resonances and the Wigner time delay, as functions of the
system size. Their equations are in fact quite similar to ours, and a
number of results differ in the two cases only because of the
differences in the details of the model studied.  Their model has two
channels per cell (particles moving to the left or right) but the
particles can have a wide range of energies, and in some instances the
high-energy limit is considered.  The similarities with the work of
Godoy {\em et al.} occur because the version of the quantum multibaker
model considered in the present paper is the simplest possible, while
more complicated versions, to be considered in further papers have no
direct counterparts in their work.

Despite the similarities between our work and that of other authors,
the focus of the work mentioned above generally differs from ours. We
are particularly interested in comparing and contrasting quantum and
classical multibaker maps, and in generalizing the QMB in a number of
directions.  These include an examination of the behavior of the QMB
for smaller values of $h$ including the semi-classical limit, and
looking for traces, if any, of the chaotic classical behavior in the
quantum version. The present paper is designed to identify important
quantum phenomena that differ from those of the classical multibaker
at large $h$, but which are expected to approach the classical results
as the Planck's constant, $h=1/N$, tends to zero.

The plan of the paper is as follows: In Section~\ref{sec:class_mb} we
will define the classical version of the model and study the evolution
of piecewise constant probability densities.  In
Section~\ref{sec:quant_mb} we quantize the multibaker map using Weyl
quantization. There we will define the uniform and random QMB's and
obtain expressions for the time dependent propagator appearing in the
discrete time version of Schr\"odinger's equation. In
Section~\ref{sec:regular} we describe the behavior of the uniform QMB
for $h=1/2$. We find the eigenstates for both closed and open systems,
as well as the steady state solutions for systems with particle
reservoirs at their boundaries. We then consider the transport
properties of particles in these chains. Of particular interest in
this connection is our finding that for open chains of uniform
multibakers, and with absorbing boundary conditions, the escape of
particles from the chain is {\em sub-diffusive} despite the ballistic
transport of particles from the interior of the chain to its
boundaries. We then turn to a brief discussion of the properties of
random multibakers and show that the assumption of random phases leads
to localized wavefunctions with very different properties from the
uniform case. Our results are summarized and discussed in Section
~\ref{sec:summary}.

\section{The classical multibaker map}
\label{sec:class_mb}

The classical multibaker map provides a reversible, deterministic
realization of a one-dimensional random walk. It is the simplest
area-preserving, deterministic model for diffusion of a particle on a
one dimensional lattice, whereby the particle makes steps either to
the right or left at equally spaced time intervals.  The multibaker
map can be adjusted for any set of step probabilities, $p,q=1-p, 0\leq
p,q\leq 1$, where $p$ is the probability of making a step to the
right. The classical multibaker map is based upon the usual baker's
map, ${{B}}$ on the unit square, $(0 \leq x,y < 1)$, defined by
\begin{equation}
  {{B}}(x,y) =
  \left\{
    \begin{array}[c]{ll}
      (x/p, py), &\quad {\rm for} \; 0 \leq x < p, \\
      ((x-p)/q,  p+ qy), &\quad {\rm for}\; p \leq x < 1.
    \end{array}
  \right.
  \label{eq:b1}
\end{equation}
The multibaker map is constructed by taking a linear chain of $L$
adjacent unit squares, labeled by the index $n, n=0,\ldots,L-1$, such
that any point on the chain is labeled by the three quantities,
$n,x,y$ with $0\leq x,y <1$. Then the action of the map, ${{M}}$ on
any point is obtained by combining a baker's map with translation of
each rectangle to the right or left, as given by
\begin{eqnarray}
  {{M}}(n,x,y) &=&
  (n+1, x/p, py), \, {\rm for} \; 0 \leq x < p, \label{eq:b2} \\
  & = &   (n-1, (x-p)/q, p+ qy),\,  {\rm for}\; p \leq x < 1. \nonumber
\end{eqnarray}
This arrangement has the property that there is a probability $p$ of
choosing a point which moves one square to the right, and probability
$q$ of choosing a point which moves one square to the left. To
complete the specification of the map, one must append boundary
conditions to the transformation given by Eq.~(\ref{eq:b2}). Such
conditions may include periodic, or absorbing boundary conditions, or
one might specify that the ends of the chain are connected to
reservoirs which maintain a constant density of points at the
boundaries.  As a chaotic system, the multibaker map is a measure
preserving map with positive and negative Lyapunov exponents, given by
$\lambda_{\pm} =\pm[p\ln(1/p)+ q \ln(1/q)]$.  This map has been used
to study the properties of deterministic diffusion in a chaotic
system, studies of the connection between diffusion coefficients and
Lyapunov exponents for an open chain, a study of entropy production in
the relaxation to a uniform equilibrium state, and has been extended
to provide simple models for
viscous and heat flows as well~\cite{Gaspard98sa,TasakiG00,TasakiG99,%
gilbert00s,tasaki95s,gaspard92s,gilbert99s,VollmerTM00,%
MatyasTV00,RondoniTV00,VollmerTB98,BreymannTV98,%
VollmerTB97,BreymannTV96,tel00}.

The classical version of the quantum multibaker considered here was
discussed in~\cite{tasaki95s}.  We consider here a simple form of this
classical model constructed to be a classical version of the $h=1/2$
quantum system.  We will study the evolution of probability densities
integrated along the stable direction ($y$) and piecewise constant on
two halves of every multibaker cell (along the unstable direction).
This space of densities is $2L$ dimensional. Therefore, the evolution
operator for this class of probability densities has the same
dimension as the quantum multibaker propagator considered in
Section~\ref{sec:quant_mb}.

\subsection{Closed, periodic case}
\label{sec:discrete.closed}

We consider the classical evolution of phase space densities under the
dynamics given by Eq.~(\ref{eq:b2}) with $p=1/2$. Since the quantum
version will describe probability amplitudes in either space or
momentum, the classical counterparts are obtained by projecting the
classical densities along the $x$ or $y$-axes, respectively.  We
restrict our attention to probability densities projected onto the
unstable $x$-direction and we take them to be constant on intervals
$0\leq x <1/2$, $1/2 \leq x < 1$, $n=\mathrm{const}$, to mimic the
$h=1/2$ quantum case. Then the projected distribution is
\begin{eqnarray}
  \label{eq:4.1}
  \rho(n,x,t) & := &\int_0^1 \!\! \rho(n,x,y,t) \,dy \nonumber\\
   & = &  \left\{
    \begin{array}[c]{ll}
      \rho_L(n,t), &\quad {\rm for} \; 0 \leq x <1/2, \\
      \rho_R(n,t), &\quad {\rm for}\; 1/2 \leq x < 1,
    \end{array}
  \right.
\end{eqnarray}
and it satisfies a Frobenius-Perron equation given by
\begin{equation}
  \label{eq:4.2}
  \rho_{L,R}(n,t+1) = \frac{1}{2}
  [\rho_L(n-1,t)+\rho_R(n+1,t)],
\end{equation}
with periodic boundary conditions $\rho_{L,R}(n+L, t) =
\rho_{L,R}(n,t)$.  Since this equation is linear we may suppose that
$\rho_{L,R}(n,t)$ represent the deviations from a uniform equilibrium
state, and may take both positive and negative values.  An eigenstate
of the right hand side of Eq. (\ref{eq:4.2}) satisfies
\begin{equation}
  \label{eq:4.3}
  \lambda \rho_L(n) = \lambda \rho_R(n) = \frac{1}{2}
  [\rho_L(n-1)+\rho_R(n+1)]. 
\end{equation}
It follows that either $\lambda=0$ or $\rho_L(n) = \rho_R(n)$.
Clearly, the $L$ vectors of the form
\begin{eqnarray*}
  \label{eq:4.4}
  \rho_L(k-1)& = & -\rho_R(k+1)\neq0, \\
  \rho_L(n\neq k-1) & = & \rho_R(n\neq k+1)=0,
\end{eqnarray*}
belong to the kernel, $\lambda=0$, of the classical discrete
multibaker.  For the case where $\lambda\neq 0$ we look for solutions
of the form \( \rho_{L,R}(n) = A e^{in\theta}\) and Eq. (\ref{eq:4.2})
leads to \( \lambda = \cos \theta.\) The general solution is, then, \(
\rho_{L,R}(n) = A_1 \cos (\theta n) + A_2 \sin(\theta n),\) where
periodic boundary conditions lead to $\theta = 2 k \pi/L,$ and the
normalized eigenstates can readily be determined. For odd $L=2M+1$ we
have the $L$ solutions
\begin{enumerate}
\item $M$ solutions of the form $\rho_{L,R}(n) = %\frac{1}{\sqrt{L}}
  A \cos\frac{2k \pi n}{L}$, $\lambda = \cos \frac{2k \pi}{L}$;
  $k=1,\dots,\frac{L-1}{2}$; 
\item $M$ solutions of the form $\rho_{L,R}(n) = %\frac{1}{\sqrt{L}}
  A \sin\frac{2k \pi n}{L}$, $\lambda = \cos \frac{2k \pi}{L}$;
  $k=1,\dots,\frac{L-1}{2}$;
\item 1 solution $\rho_{L,R}(n) = %\frac{1}{\sqrt{2L}}
  A $, $\lambda = 1$; $k=0$. 
\end{enumerate}
For even $L=2M$ we have the $L$ solutions
\begin{enumerate}
\item $M-1$ solutions of the form $\rho_{L,R}(n) = %\frac{1}{\sqrt{L}}
  A \cos\frac{2k \pi n}{L}$, $\lambda = \cos \frac{2k \pi}{L}$;
  $k=1,\dots,\frac{L}{2}-1$;
\item $M-1$ solutions of the form $\rho_{L,R}(n) = %\frac{1}{\sqrt{L}}
  A \sin\frac{2k \pi n}{L}$, $\lambda = \cos \frac{2k \pi}{L}$;
  $k=1,\dots,\frac{L}{2}-1$;
\item 1 solution $\rho_{L,R}(n) = %\frac{1}{\sqrt{2L}}
  A$, $\lambda = 1$; $k=0$, and,
\item 1 solution $\rho_{L,R}(n) = %\frac{(-1)^n}{\sqrt{2L}}
  (-1)^n A $, $\lambda = -1$; $k=M=L/2$.
\end{enumerate}
We see that in the odd case there is an approach to equilibrium: all
the eigenvalues have absolute value strictly less than 1, apart from
the one corresponding to the uniform distribution. The even case is
sensitive to the ``even-odd'' oscillations of the location of a point
along the chain. These oscillations can be removed by combining two
successive steps.

\subsection{Open case (absorbing boundary conditions)}
\label{sec:discrete.open}

For the open chain with absorbing boundary conditions, the dynamics
inside is the same as in the closed case and is given by
Eq.~(\ref{eq:4.2}), therefore the general solution is also given by \(
\rho_{L,R}(n) = A_1 \cos (\theta n) + A_2 \sin(\theta n).\) Absorbing
boundary conditions $\rho_{R,L}(-1) = \rho_{R,L} (L) = 0$ lead to
\begin{widetext}
\begin{eqnarray}
  \label{eq:4.10}
  \lambda A_1 & = & \frac{1}{2} [A_1 \cos \theta + A_2 \sin \theta], \\ 
  \label{eq:4.11} 
  \lambda [A_1\cos (L-1)\theta + A_2 \sin (L-1) \theta] & = & 
  \frac{1}{2} [A_1 \cos (L-2) \theta + A_2 \sin (L-2) \theta] ,
\end{eqnarray}
\end{widetext}
where $\lambda=\cos \theta$.  They have nontrivial solutions if and
only if \( \sin (L+1) \theta = 0,\) leading to \( \theta_k =
\frac{k\pi}{L+1},\) where $k=-L,\dots,-1,1,\dots,L$, and
$\lambda(-\theta)=\lambda(\theta)$. Thus finally $k=1,\dots,L$, which
gives the $L$ states of the form
\begin{equation}
  \rho_{L,R}(n) = A \sin \frac{k (n+1) \pi}{L+1}.
\end{equation}
The remaining $L$ states are in the kernel: $L-2$ are of the form
given by Eq. (\ref{eq:4.4}) with $n=1,\dots,L-2$.  Two other states
corresponding to $\lambda =0$ are \( \rho_L(L-1) = 1,\ \rho_L(k\neq
L-1)=\rho_R(k) =0, \) and \( \rho_R(0) = 1,\ \rho_R(k\neq 0)=\rho_L(k)
=0. \) Thus we easily obtain a spectral decomposition for the simple
operator treated here, with absorbing boundary conditions.

The probability of finding the particle in the system decays with the
escape rate
\begin{equation}
  \label{eq:escape1}
  \gamma := - \lim_{t\rightarrow \infty} \frac{\log P(t)}{t},
\end{equation}
where $P(t) := \sum_n \rho(n,t)$, given by the largest eigenvalue
\begin{equation}
  \gamma = -\log |\cos \frac{\pi}{L+1}| \approx \frac{\pi^2}{2 L^2},
\end{equation}
for large $L$.

\subsection{ The open, discrete multibaker with reservoirs}
\label{sec:discrete.reservoirs}

\begin{figure}[htbp]
  \centering
  \includegraphics[scale=0.5]{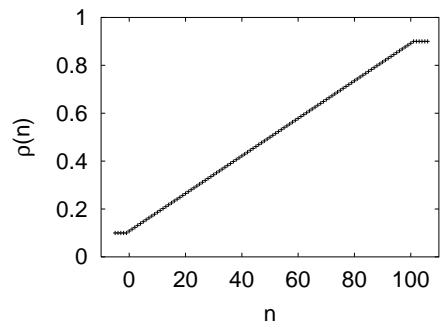}
  \caption{Linear profile of the classical discrete multibaker map
    of length $L=101$ with reservoirs ($\rho_1=0.1, \rho_2=0.9$). The
    horizontal axis range is $[-5,105]$.} 
  \label{fig:linear}
\end{figure}

Next we connect particle reservoirs to a finite chain and look
for steady state solutions. These are time invariant solutions to 
Eq. (\ref{eq:4.2}), with the boundary conditions
\begin{eqnarray}
  \rho_L(0) & = & \rho_R(0) = \frac12 [\rho_1 + \rho_R(1)],\\
  \rho_L(L-1) & = & \rho_R(L-1) = \frac12 [\rho_L(L-1) + \rho_2],
\end{eqnarray}
where $\rho_1, \rho_2$ are the incoming densities of the left and
right reservoirs, respectively.  A solution is found immediately by
observing that in the steady state $\rho_L(n)=\rho_R(n)\equiv\rho(n)$
and that Eq.~(\ref{eq:4.2}) leads to \( \rho(n+1) = 2\rho(n) -
\rho(n-1).\) A solution satisfying the boundary conditions is
therefore
\begin{equation}
  \rho(n) = \frac{\rho_2 + L \rho_1}{L+1} + \frac{n(\rho_2 -
  \rho_1)}{L+1}.
\end{equation}
This linear profile expected from the Fick's
law~\cite{gaspard92s,tasaki95s,Gaspard98sa} is shown in
Figure~\ref{fig:linear}. 

\section{Quantum multibaker map: the general model}
\label{sec:quant_mb}

In order to quantize the multibaker map, we start with the quantum
baker map, as described by Balazs, Voros \cite{BalazsV89}, and
Saraceno \cite{Saraceno90}, and then produce a quantum multibaker map
by forming a chain of unit squares, applying the quantum baker map to
each square, but transferring the new quantum states to the adjacent
squares according to the procedure used in the classical case.

The method for constructing a quantum version of the regular baker map
is as follows. We consider the $x$-direction to be the ``spatial''
direction of the system, and the $y$-direction to be the ``momentum''
direction. Then the number of quantum states $N$ in the unit square
should satisfy $N = PQ/ 2\pi\hbar$, where $Q=1$ is the spatial extent
of the unit square, and $P=1$ is the range of momenta. This leads to
the simple formula, $\hbar=1/(2\pi N)$, where $N$ is an integer. We
usually, but not always, take $N$ to be an even integer, so that one
half of the quantum states can be associated with each half of the
unit square.  One then constructs a set of $N$ ``position'' states for
a unit square, with position eigenvalues $q_j=2\pi \hbar
(j+\phi_q)=(j+\phi_q)/N,\; j=0,\ldots, N-1$ and a set of $N$
``momentum'' states with momentum eigenvalue $p_k = 2\pi \hbar
(k+\phi_p)=(k+\phi_p)/N, \; k=0, \ldots, N-1$. We require that
$0\leq\phi_{q,p}<1$. The position and momentum states are related to
each other by means of a simple Fourier representation, with $N$
terms, given by
\begin{equation}
  (G_N)_{kj} \equiv <p_k|q_j> = \frac{1}{\sqrt{N}}e^{-2\pi i (k +
  \phi_p)(j+ \phi_q)/N}. 
  \label{eq:b3}
\end{equation}
We include subscripts on $p,q$ in the notation for the Dirac matrix
element to identify the integers which are attached to the $p$ and $q$
representations.  The phases, $\phi_q,\phi_p$ are as yet unspecified.
In the literature on the quantum baker map, these phases are often
taken to be 0 (the simplest~\cite{BalazsV89}) or 1/2 (most symmetric
map~\cite{Saraceno90}).  Here we will take advantage of the
possibility to choose these phases so as to represent different
situations that may have some relevance to physical phenomena.

The time dependence of the quantum baker map is determined by
constructing a propagator for the change in the quantum states over
one time step.  This propagator consists of two parts: First one
transforms the ``left'' part of the Hilbert space (in the position
representation) into ``bottom'' subspace (in momentum representation)
and the ``right (position) part'' into the ``top (momentum) part''.
Then one uses the Fourier relation between position and momentum
states,~Eq.~(\ref{eq:b3}), to change the basis from momentum back to
position representation.  The first transformation consists of two
Fourier transforms over $N/2$-dimensional space, the other is the
inverse Fourier transform over the whole $N$-dimensional space
\begin{equation}
  {\bf{B}} = \left[ G_N^{-1} \right] \cdot
  \left[ 
    \begin{array}[c]{cc}
      G_{N/2} & 0 \\
      0 & G_{N/2}
    \end{array}    
  \right] .
\label{eq:b4}
\end{equation}
The action of ${\bf{B}}$ on a position-space wave function is
understood as follows.  We represent the position-space function as a
column vector with $N$ elements, the top $N/2$ elements referring to
the quantum states with numbers $j=0,1,\ldots, (N/2) -1$, which we
denote as the ``left'' states. The bottom $N/2$ elements having
quantum numbers $j= N/2, N/2 +1, \ldots, N-1$, are called the
``right'' states. The block diagonal matrix, with blocks $G_{N/2}$
appearing on the right hand side of Eq. (\ref{eq:b4}) transfers the
left and right spatial states to the ``bottom'' and ``top'' momentum
states, respectively, according to
\begin{equation}
  \left[
    \begin{array}[c]{c}
      \widetilde{\Psi}_b(t+1) \\ \widetilde{\Psi}_t(t+1)
    \end{array}
  \right] = 
  \left[
    \begin{array}[c]{cc}
      G_{N/2}(\phi_q,\phi_p) & 0\\ 
      0 & G_{N/2}(\phi_q,\phi_p)
    \end{array}
  \right]\cdot
  \left[
    \begin{array}[c]{c}
      \Psi_l(t) \\ \Psi_r(t)
    \end{array}
  \right].
\label{eq:b6}
\end{equation}
This operation defines the quantum baker map. However, we are left
with a quantum state in the momentum representation. We now change the
momentum state representation into a spatial state by means of the
matrix $G_{N}^{-1}$, as in Eq. (\ref{eq:b3}).

Finally we can construct a {\em quantum multibaker} map (QMB) by
considering a chain of unit squares, each taken to be an individual
quantum system, but which exchange quantum states according to the
rules of the quantum baker with an interlacing process formed in
analogy with the classical multibaker map Eq. (\ref{eq:b2}). That is,
the position space functions at site $n$ are transformed to momentum
space functions at sites $n\pm 1$, according to the rule
\begin{equation}
  \left[
    \begin{array}[c]{c}
      \widetilde{\Psi}_b(n+1,t+1) \\ \widetilde{\Psi}_t(n-1,t+1)
    \end{array}
  \right] = 
  \left[
    \begin{array}[c]{cc}
      G_{N/2}(n) & 0\\ 
      0& G_{N/2}(n)
    \end{array}
  \right]\cdot
  \left[
    \begin{array}[c]{c}
      \Psi_l(n,t) \\ \Psi_r(n,t)
    \end{array}
  \right].
\label{eq:b8}
\end{equation}
Here we can allow for the phases $\phi_{q,p}(n)$ to vary from one
cell, denoted by $n$, to the next, and we incorporate them in the
transformation operators $G_N(n) \equiv G_N(\phi_q(n),\phi_p(n))$ at
that site. After this transformation is carried out, we change from
the momentum to the position representation at each site according to
the same rule as in an ordinary quantum baker map, that is
\begin{equation}
  \left[
    \begin{array}[c]{c}
      \Psi_l(n,t+1) \\ \Psi_r(n,t+1)
    \end{array}
  \right] = 
  G_N^{-1}(n) \cdot
  \left[
    \begin{array}[c]{c}
      \widetilde{\Psi}_b(n,t+1) \\ \widetilde{\Psi}_t(n,t+1)
    \end{array}
  \right] .
\label{eq:b9}
\end{equation}
\begin{widetext}
{\em Thus in the position representation the quantum multibaker map is
  given by}
\begin{equation}
  \left[
    \begin{array}[c]{c}
      \Psi_l(n,t+1) \\ \Psi_r(n,t+1)
    \end{array}
  \right]  
  =
  G_N^{-1}(n) \cdot
  \left[
    \begin{array}[c]{cc}
      G_{N/2}(n-1) & 0\\ 
      0 & G_{N/2}(n+1)
    \end{array}
  \right]\cdot
  \left[
    \begin{array}[c]{c}
      \Psi_l(n-1,t) \\ \Psi_r(n+1,t)
    \end{array}
  \right].  \label{eq:multidyn}
\end{equation} 
Explicitly, we have
\begin{eqnarray}
  \Psi_l(n,t+1) & = & [G_N^{-1}(n)]_{l,b}\cdot G_{N/2}(n-1)\cdot
  \Psi_l(n-1,t) 
  + [G_N^{-1}(n)]_{l,t}\cdot G_{N/2}(n+1)\cdot \Psi_r(n+1,t), \label{c1}\\
  \Psi_r(n,t+1) & = & [G_N^{-1}(n)]_{r,b}\cdot G_{N/2}(n-1)\cdot
  \Psi_l(n-1,t) 
  + [G_N^{-1}(n)]_{r,t}\cdot G_{N/2}(n+1)\cdot \Psi_r(n+1,t).
  \label{c2}
\end{eqnarray}
\end{widetext}
Here, in an obvious notation, the matrices
$[G_{N}^{-1}]_{\alpha,\beta}$ are $N/2\times N/2$ block sub-matrices
that comprise $G_N^{-1}$. The general case can be treated numerically,
of course, once the phases are specified. It is of interest to
consider the special case $N=2$, since much of the work can be done
using simple analytical methods, and since this case corresponds to
the largest possible value for Planck's constant, namely, $h=1/2$.
This is the case we study here.

The local dynamics are characterized by the two phases, $\phi_q,
\phi_p$, which parameterize the Weyl quantizations of the baker map.
If we take the same pair of phases at each site we obtain the {\em
  uniform\/} model. If we choose them randomly from some distribution
at each of the sites, we get the {\em random\/} model. In this paper,
when we treat the random model we will assume that the phases are
chosen according to a uniform distribution on the unit
circle~\footnote{P. Gaspard has suggested that the phases used here
  may be thought of as resulting from vector potentials produced by
  thin solenoids, which modify the phases of the wave-functions
  through Bohm-Aharonov loops.}.

A complete specification of the model is obtained by adding the
boundary conditions to the above equations. In this work we restrict
our attention to the closed case (with periodic boundary conditions),
and open cases (with either absorbing boundary conditions or with
``particle'' reservoirs at the ends of the chain).

\section{Uniform quantum multibaker}
\label{sec:regular}

The uniform quantum multibaker is characterized by a set of phases
$\phi_q,\phi_p$ that are independent of the site index, that is, they
are the same for each of the transformation matrices generating the
map, as described in Eq. (\ref{eq:multidyn}). This makes transport in
the uniform multibaker chain similar in many respects to transport in
a one-dimensional periodic solid. Here we solve this model for time
dependent and stationary quantum states with appropriate boundary
conditions: the closed, periodic chain; the open chain with absorbing
boundary conditions; and the open chain attached to leads at each end,
producing a stationary, non-equilibrium state. We begin with the
periodic chain.

\subsection{Closed, periodic case}
\label{sec:regular.closed}

We consider the periodic, uniform multibaker chain, with $L$ sites and
$N=2$. The equation connecting the quantum states at time $t+1$ to
those at time $t$ is
\begin{eqnarray*}
  \Psi_l(n,t+1) & = & f_0 (g_{00} \Psi_l(n-1,t) +  g_{01}
  \Psi_r(n+1,t)), 
  \\ 
  \Psi_r(n,t+1) & = & f_0 (g_{10} \Psi_l(n-1,t) +  g_{11}
  \Psi_r(n+1,t)). 
\end{eqnarray*}
with ($0\leq \phi_q, \phi_p <1$), and $f_0, g_{kl}$ given by
\begin{equation}
  \begin{array}[c]{rcccl}
   \displaystyle f_0 & = & \displaystyle(G_1(\phi_q,\phi_p))_{00} & = &
   \displaystyle e^{-i2\pi\phi_q\phi_p}, \\
   \displaystyle g_{kl} & = &
   \displaystyle(G^{-1}_2(\phi_q,\phi_p))_{kl} & = & \displaystyle \frac{1}{\sqrt{2}} 
    e^{i\pi(k+\phi_q)(l+\phi_p)}.    
  \end{array}
  \label{eq:b10}
\end{equation}
Since the system is periodic, Bloch's theorem guarantees the existence of
eigenstates of the form
\begin{equation}
  \label{eq:Bloch}
  \Psi_{r,l}(n) =  A_{r,l} \chi^n = A_{r,l} e^{in\theta}.
\end{equation}
Periodic boundary conditions, $\Psi(L)=\Psi(0)$, imply that $\theta= 2
k \pi/L, \,\, k=0,\ldots, L-1$.  Clearly, $\lambda$ is an eigenvalue
of the quantum multibaker propagator if and only if
\begin{equation}
  \left |
    \begin{array}[c]{cc}
      g_{00}f_0 e^{-i\theta} -\lambda&
      g_{01}f_0 e^{i\theta} \\
      g_{10}f_0 e^{-i\theta} & 
      g_{11}f_0 e^{i\theta} -\lambda
    \end{array}
  \right | = 0.
\label{eq:c3}
\end{equation}
Using the notation
\begin{equation}
  \begin{array}[c]{rcl}
  \alpha & = & (1+\phi_q+\phi_p)\pi/2, \\
  \beta & = & (1+\phi_q+\phi_p-2\phi_q\phi_p)\pi/2 = \alpha-\pi \phi_q
  \phi_p, 
  \end{array}
\label{eq:alphabeta}
\end{equation}
we find that  
\begin{equation}
  \lambda = \frac{e^{i\beta}}{\sqrt{2}}[\cos (\theta+\alpha) \pm
  i \sqrt{1 + \sin^2 (\theta+\alpha)}].
\label{eq:c4}
\end{equation}
Note that~(\ref{eq:c3}) can also be written as
\begin{eqnarray}
  \lambda/e^{i\beta} + e^{i\beta}/\lambda & = & \frac{1}{\sqrt{2}}
  \left[e^{i(\alpha+\theta)} + e^{-i(\alpha+\theta)}\right], \nonumber
  \\
  v+\frac{1}{v} & = &\frac{1}{\sqrt{2}} \left[u + \frac{1}{u}\right]
  ,\label{eq:uandvtemp2} 
\end{eqnarray}
where $u=\chi e^{i\alpha}$, $v=\lambda/e^{i\beta}$.  Since $\theta$ is
real it follows that $|\lambda|^2=1$, so $\lambda=e^{i\gamma}$ and \(
\gamma -\beta \in [\pi/4,3\pi/4]\cup[5\pi/4,7\pi/4].  \) Therefore the
``quasi-energies'', $\gamma$, lie in two bands of length $\pi/2$
symmetric with respect to the center of the unit circle. The exact
location depends upon the phases $\phi_q,\phi_p$.  Making use of the
boundary conditions we obtain the eigenvalues of the closed multibaker
map
\begin{equation}
  \lambda_{\pm,k} = \frac{e^{i\beta}}{\sqrt{2}}\left[\cos
  (\alpha+2k\pi/L) \pm i \sqrt{1 + \sin^2 (\alpha+2k\pi/L)}\right],
\end{equation}
with $\alpha$ and $\beta$ given by Eq.~(\ref{eq:alphabeta}).  The
corresponding eigenstates are given by Eq.~(\ref{eq:Bloch}) with the
constants connected by $A_r = A_l[\sin(\alpha+2k\pi/L)\mp\sqrt{1 +
  \sin^2 (\alpha+2k\pi/L)} ] e^{i (\pi(\phi_p-\phi_q)/2-2k\pi/L)}$.

\begin{figure}[htbp]
  \centering
  \includegraphics[scale=0.5]{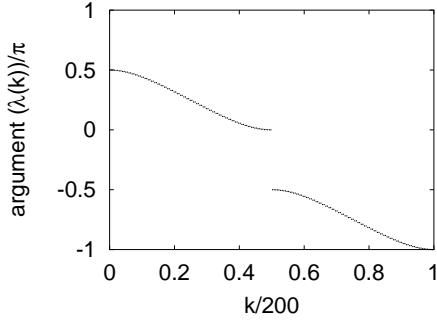}
  \caption{Eigenspectrum of the closed uniform quantum multibaker  for
  the chain of length $L=101$ cells with periodic boundary
  conditions. Phases are $\phi_q=\phi_p = 1/2, \alpha=\pi,
  \beta=3\pi/4$. } 
  \label{fig:reg.band}
\end{figure}

When $\alpha$ is an integer multiple of $\pi/L$ the spectrum is doubly
degenerate. This non-generic case happens for instance for the most
common choices of phases ($\phi_q=\phi_p=0$ or $1/2$).  The
quasi-energy spectrum of the closed uniform multibaker for the phases
$\phi_q=\phi_p=1/2$ is shown in Figure~\ref{fig:reg.band}.

\subsection{Open case: absorbing boundary conditions}
\label{sec:regular.open}

Next we consider the uniform quantum multibaker, still for $N=2$, but
with open boundaries. In the classical case, open boundaries are
important for the application of the escape-rate formalism of Gaspard
and Nicolis~\cite{gaspard90s} which relates the rate of decay of the
initial number of particles on a large, open chain to the diffusion
coefficient, and then to the Lyapunov exponents and the
Kolmogorov-Sinai entropy of trajectories on a fractal repeller, {\it
  i.e.\/} the set of initial points for trajectories that never leave
the chain~\cite{Gaspard98sa,Dorfman99s}. It is of some interest, then,
to contrast the classical and quantum cases.

We take the multibaker dynamics given by~(\ref{eq:multidyn}) in the
cells $n= 1, 2, \dots, L-2$. At the boundary cells we allow the
probability density to escape from the right half cell for $n=0$, and
from the left half cell for $n=L-1$, and nothing enters the system
from the outside. The latter condition requires
\begin{eqnarray}
  \Psi_b(0,t) & = & 0, \label{eq:b24}\\
  \Psi_t(L-1,t) & = & 0. \label{eq:b25}
\end{eqnarray}

Due to the escape of probability density, the eigenvalues that
determine the time dependence of the probability density in each cell
move to the interior of the unit circle.
A simple proof of this fact is given in
Appendix~\ref{sec:app.eigenv.open}. We show there also that the kernel
is two-dimensional.

\begin{figure}[htbp]
  \centering
  \includegraphics[scale=0.5]{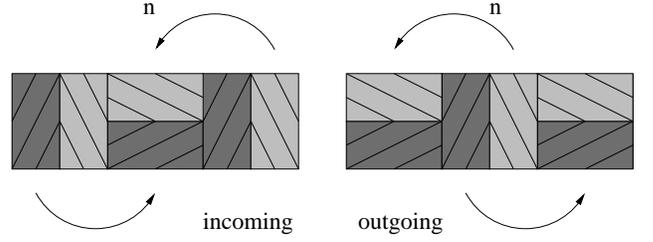}
  \caption{Scattering from one cell.}
  \label{fig:scattercell}
\end{figure}

To determine the non-zero eigenvalues $0 < |\lambda| <1$ of the open
chain, we first write the eigenvalue equation in the momentum
representation.  Then every eigenstate $\Psi$ satisfies the equation
\begin{equation}
  \left[
    \begin{array}[c]{c}
     \widetilde \Psi_b(n+1) \\ \widetilde\Psi_t(n-1)
    \end{array}
  \right] = 
  \left[
    \begin{array}[c]{cc}
      f_0/\lambda & 0\\
      0 & f_0/\lambda
    \end{array}
  \right]\cdot
  G_2^{-1}(\phi_q,\phi_p)\cdot
  \left[
    \begin{array}[c]{c}
      \widetilde\Psi_b(n) \\ \widetilde\Psi_t(n)
    \end{array}
  \right]  .
  \label{eq:d1}
\end{equation}
Viewed in terms of the ``top'' and ``bottom'' states, we see that the
solution of Eq. (\ref{eq:d1}) can be neatly formulated as a scattering
problem (see Figure~\ref{fig:scattercell}), 
where the incoming waves are $\widetilde\Psi_b(n)$ and
$\widetilde\Psi_t(n)$, and the outgoing waves are
$\widetilde\Psi_b(n+1)$ and $\widetilde\Psi_t(n-1)$ with a one-cell
scattering $S$-matrix, and a one-cell transfer $T$-matrix. Those are
defined, respectively, by the relations
\begin{equation}
  \left[
    \begin{array}[c]{c}
      \widetilde\Psi_t(n-1) \\ \widetilde\Psi_b(n+1)
    \end{array}
  \right] = 
  S \cdot
  \left[
    \begin{array}[c]{c}
      \widetilde\Psi_b(n) \\ \widetilde\Psi_t(n)
    \end{array}
  \right],
\label{sm}
\end{equation}
and
\begin{equation}
  \left[
    \begin{array}[c]{c}
      \widetilde\Psi_b(n+1) \\ \widetilde\Psi_t(n)
    \end{array}
  \right] = 
  T \cdot
  \left[
    \begin{array}[c]{c}
      \widetilde\Psi_b(n) \\ \widetilde\Psi_t(n-1)
    \end{array}
  \right]  .
\end{equation}
Explicitly, the $S$-matrix is, for the uniform multibaker, given
\begin{equation}
  S = 
  \left[
    \begin{array}[c]{cc}
      \frac{1}{\sqrt{2}} e^{i\pi \phi_p(1-\phi_q)}/\lambda & 
      \frac{1}{\sqrt{2}} e^{i\pi (1+\phi_q+\phi_p-\phi_q\phi_p)}/\lambda \\
      \frac{1}{\sqrt{2}} e^{-i\pi \phi_q \phi_p}/\lambda & 
      \frac{1}{\sqrt{2}} e^{i\pi \phi_q(1-\phi_p)}/\lambda
    \end{array}
  \right],
\end{equation}
and the $T$-matrix is 
\begin{equation}
  T = 
  \left[
    \begin{array}[c]{cc}
      \sqrt{2} e^{-i\pi \phi_q\phi_p}/\lambda & 
      - e^{-i\pi \phi_p} \\
      e^{-i\pi \phi_q } & 
      \sqrt{2} e^{-i\pi (1+\phi_q+\phi_p-\phi_q\phi_p)}\lambda
    \end{array}
  \right].
\end{equation}
 
\begin{figure}[htbp]
  \centering
  \includegraphics[scale=0.65]{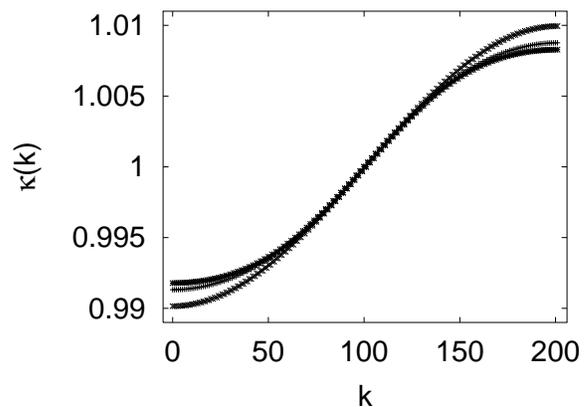} 
  \caption{The first and the fifth order approximate solutions and the
  numerical solutions of Eq.~(\ref{eq:pert}) for $L=101$. We took here
  $\delta=1$. The numerical solution is the middle curve. Fifth order
  approximate solution is the one closer to the numerical solution.} 
  \label{fig:series}
\end{figure}

We find it convenient to use the transfer operators, $T$, to carry out
the determination of the eigenvalues, $\lambda$, governing the rate of
decay for an open system. To do this we first use the transfer
operators to relate the quantum states at one end of the chain to the
states at the other end, and then use the open, absorbing boundary
conditions to obtain an explicit equation for $\lambda$.  First, the
states at the two ends of the chain are related by
\begin{equation}
  \left[
    \begin{array}[c]{c}
      \widetilde\Psi_b(L-1) \\ \widetilde\Psi_t(L-2)
    \end{array}
  \right] = 
  T^{L-2}\cdot
  \left[
    \begin{array}[c]{c}
      \widetilde\Psi_b(1) \\ \widetilde\Psi_t(0)
    \end{array}
  \right]  .
  \label{eq:36}
\end{equation}
To use the boundary conditions, we first look at cell $L-1$. We note
that in the open multibaker $\widetilde\Psi_t(L-1)=0$. Then using
Eq. (\ref{eq:d1}), we obtain
\(  \widetilde\Psi_t(L-2) = 
  (f_0/\lambda)\, [g_{10} \widetilde\Psi_b(L-1)+g_{11}\widetilde\Psi_t(L-1)], \)
and
\(  \widetilde\Psi_t(L-1) = [(\lambda/f_0) \widetilde\Psi_t(L-2) -
g_{10}\widetilde\Psi_b(L-1)] / g_{11}.\) 
Thus
\begin{equation}
  0 =
  \left[
  \begin{array}[c]{cc}
    -g_{10} & (\lambda/f_0)
  \end{array}
  \right]\cdot
  \left[
    \begin{array}[c]{c}
      \widetilde\Psi_b(L-1) \\
      \widetilde\Psi_t(L-2)
    \end{array}
  \right].
\end{equation}
Using Eq.~(\ref{sm}) and~(\ref{eq:36}), we easily find that \(
\widetilde\Psi_b(1) = (f_0 g_{01}/\lambda) \widetilde\Psi_t(0). \)
Thus the equation that determines the decay rates is
\begin{equation}
  0 =
  \left[
  \begin{array}[c]{cc}
    -f_0 g_{10} & \lambda
  \end{array}
  \right]\cdot
  T^{L-2}\cdot
  \left[
    \begin{array}[c]{c}
      f_0 g_{01} \\
      \lambda
    \end{array}
  \right], 
\label{eq:d2}
\end{equation}
where a scalar product of matrices is to be taken as indicated.  To
get a useful form for this equation we need to find the eigenvalues of
the transfer matrix $T$.  We denote the eigenvalues of $T$ by $\chi_+,
\chi_{-}$, which are obtained as solutions of the quadratic equation
\begin{equation}
  \chi e^{i\alpha} + \frac{1}{\chi e^{i\alpha}}  =  \sqrt{2}
  [\lambda/e^{i\beta} + e^{i\beta}/\lambda] ,
\end{equation}
where $\alpha, \beta$ are given by Eq.~(\ref{eq:alphabeta}). Using, as
before, the notation $u=\chi e^{i\alpha}$, $v=\lambda/e^{i\beta}$ we
obtain the same formal relation between $u$ and $v$ as in the periodic
case, Eq. (\ref{eq:uandvtemp2}), {\it i.e.}
\begin{equation}
  \label{eq:uandv}
  v + \frac{1}{v} =  \frac{1}{\sqrt{2}}\left[u + \frac{1}{u}\right],
\end{equation}
The two solutions $u_+, u_-$ satisfy $u_+ u_- = 1$, and $u_+ + u_- =
\sqrt{2}[v + 1/v]$.  Since $|v|=|\lambda|<1$, it follows that $u_+,
u_-$ do not lie on the unit circle. In particular, they must be
different and so the matrix $T$ is non-degenerate. We take
$|u_+|>1>|u_-|$ to define them uniquely, and use $u_\pm = \chi_\pm
e^{i\alpha}$. If we set $u_\pm=e^{\pm i\kappa}$,
and then solve for $v$ we obtain
\begin{equation}
  \label{eq:vofu}
  v_\pm = \frac{1}{\sqrt{2}} [\cos\kappa \pm i  \sqrt{1+\sin^2\kappa}].
\end{equation}
Interesting solutions are those where $\kappa$ is not purely real,
that is, $\kappa\in\mathbb{C}\setminus\mathbb{R}$.  We next use a
simple identity for the $L$-th power of non-degenerate matrix $T$,
given by
\begin{eqnarray*}
  T^L & = &\frac{\chi_+^L-\chi_-^L}{\chi_+-\chi_-}T
  -\frac{\chi_-\chi_+^L-\chi_+\chi_-^L}{\chi_+-\chi_-}I \\
  &=& \frac{e^{-i \alpha (L-1)}}{\sin \kappa} [\sin (L\kappa) \,T - \sin
  ((L-1)\kappa) \, I] ,
\end{eqnarray*}
to write
Eq.~(\ref{eq:d2}) in the form
\begin{equation}
  \label{eq:ul}
  \left[2v^2+2+\frac{1}{v^{2}}\right] \sin (L-2)\kappa =
   \left[\sqrt{2}v+\frac{1}{\sqrt{2}v}\right] \sin(L-3)\kappa ,
\end{equation}
where $v$ is one of $v_\pm$.  With the help of~(\ref{eq:uandv})
and~(\ref{eq:vofu}) we can reduce~(\ref{eq:ul}) to
\begin{equation}
  \sin L\kappa +\sin \kappa \cos (L-1)\kappa + i\epsilon
  \sqrt{1+\sin^2\kappa}\sin (L-1)\kappa = 0,
\end{equation}
which can be further reduced to
\begin{equation}
  \label{eq:u2l}
  u^{2L} -1 -2\sin\kappa [\sin\kappa-\epsilon \sqrt{1+\sin^2\kappa}] = 0.
\end{equation}
In the above equations $\epsilon=\pm 1$ corresponds to the sign
in~(\ref{eq:vofu}). Clearly, we can get all of the possible solutions
multiplying Eq.~(\ref{eq:u2l}) for two different signs. This leads to
a very simple equation
\begin{equation}
  \label{eq:sin2}
  \sin^2 L\kappa + \sin^2 \kappa = 0.
\end{equation}
The only real solutions of this equation are $\kappa = k \pi, \,
k\in\mathbb{Z}$, but, as mentioned above, they must be discarded.  If
we write Eq.~(\ref{eq:sin2}) as
\begin{equation}
  \label{eq:pert}
  \sin L\kappa =i \delta \sin \kappa,
\end{equation}
where $\delta=\pm 1$, we can treat it as a ``perturbation'' in
$\delta$ of equation $\sin L\kappa = 0$~\cite{bender78m.pert}.  Thus
we can obtain the solutions of interest by expanding $\kappa$ in
powers of $\delta$ about the values $\kappa = k\pi/L,\, k= 1,\ldots,
L-1$,~\footnote{Remaining values $k=L+1,\ldots,2L-1$ lead to $u
  \rightarrow 1/u$ which gives the same $v$ and thus they are
  discarded.} and then at the end, setting $\delta =\pm 1$. This
approach gives results which quickly converge numerically, for all
allowed values of $k$. To apply this procedure it is convenient to
rewrite Eq.~(\ref{eq:pert}) in a polynomial representation,
\begin{equation}
  \label{eq:pert1}
  u^{2L}-1-\delta iu^L(u-1/u) = 0 .
\end{equation}
Then, by taking 
\(
u = \exp(i(k \pi/L+\delta a_1 + \delta^{2} a_2 +\dots)),
\)
one can determine the coefficients $a_i$, and check the convergence of
the series numerically. Figure~\ref{fig:series}
shows the absolute values of the approximate solutions (in the first
and fifth order) and the numerical solutions for $L=101$ sorted
according to increasing amplitude.  The first few coefficients in the
expansion of $u$ are
\begin{eqnarray*}
  a_1 & = & -b/L, \\
  a_2 & = & -iab/L^2, \\
  a_3 & = & b^3/6L + b(2-3b^2)/2L^3,\\
  a_4 & = & i(2 ab^3/3L^2 - ab(8b^2-3)/3L^4),
\end{eqnarray*}
where
\begin{equation}
  \label{eq:aandb}
  a = \cos(k \pi/L), \qquad b = \sin(k \pi/L).
\end{equation}
Numerical studies show rapid convergence of amplitudes and slower
convergence of phases.

\begin{figure}[htbp]
  \label{fig:abv}
  \centering
    \includegraphics[scale=0.65]{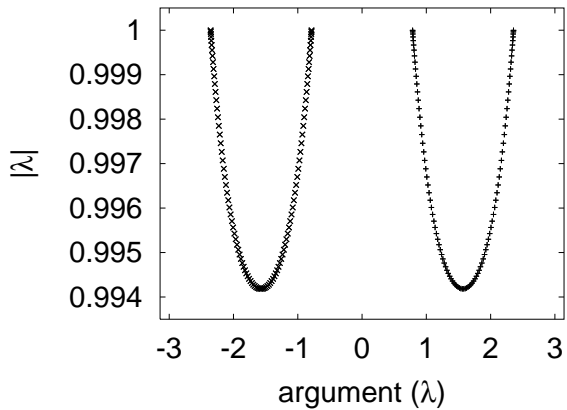} 
  \caption{The amplitude versus argument of $\lambda$ in the fourth order of
    approximation
    for the chain of length $L=101$.
    }
\end{figure}

Next we calculate the approximate eigenvalues of the open quantum
multibaker. Using $\lambda = e^{i \beta} v$, keeping $|v|<1$
solutions, to second order in $\delta$ we obtain
\begin{widetext}
\begin{equation}
  \label{eq:approx_v}
  \lambda = \frac{e^{i\beta}}{\sqrt{2}} \left( a+i \epsilon\,\sqrt
    {1+{b}^{2}} \right)\,   
  \exp\left\{-\frac{b^2}{L\,\sqrt{1+b^2}}\right\}
  \exp\left\{-\frac{i\epsilon a b^2 \left( 2\,b^2+3 \right)}{2
      L^2 \left( 1+b^2 \right) ^{3/2}}\right\} , 
\end{equation}
\end{widetext}
where $a,b$ are given by~(\ref{eq:aandb}), while $\epsilon=\pm1$
enumerates the solutions.  The non-exponential factor on the right
hand side is the unperturbed solution.  Figure~\ref{fig:abv} shows the
absolute value of $v$ (in the fourth order approximation).

The general result and the numerical studies suggest that the value of
$v$ that is the nearest to the unit circle occurs when $k$ takes on
one of the four values, $k\in\{1,L-1,L+1,2L-1\}$. To get the leading
term in the large-size limit we find it convenient to use an
alternative expansion of Eq.~(\ref{eq:pert}), in powers of $L^{-1}$.
While this leads to an asymptotic solution for $v$ which quickly
diverges for most $k$, it gives us the correct leading order behavior
for $k\ll L$.  Then the expansion of the solution of Eq.
(\ref{eq:pert}) in powers of $1/L$, as
$\kappa_k=k\pi/L+b_2/L^2+b_3/L^3+\dots$ for small $k$ yields
\begin{equation}
  \kappa_k \approx k\pi(1/L + i \delta/L^2 -1/L^3 - i\delta (1+k^2
  \pi^2/3)/L^4 + \dots), 
\end{equation}
which gives the asymptotic formula for $v(k)$
\begin{eqnarray}
  v(k) & = & \frac{1}{\sqrt{2}} [\cos\kappa_k \pm i \sqrt{1+\sin^2(\kappa_k)}] \\
  & \approx & \exp \left[\pm i\left(\frac{\pi}{4} +
  \frac{k^2\pi^2}{2L^2}\right)\right] \exp\frac{-k^2 
  \pi^2}{L^3}.
\end{eqnarray}
The escape of probability density from an open system asymptotically
is dominated by the eigenvalue closest to the unit circle.  Therefore
the escape rate
\begin{equation}
  \label{eq:escape}
  \gamma := - \lim_{t\rightarrow \infty} \frac{\log P(t)}{t}
\end{equation}
of the uniform quantum multibaker map is obtained from the eigenvalue
corresponding to $k=1$:
\begin{equation}
  \gamma = -\log |v(1)|^2 \approx \frac{2 \pi^2}{L^3}.
\end{equation}

This result means that even though the motion {\em inside\/} the
quantum multibaker is {\em faster\/} (ballistic) than in the
corresponding classical system (diffusive), the {\em effusion\/}
(decay of probability density) is {\em slower} than that for the
corresponding classical system, (\cite{Gaspard98sa}; see also
Section~\ref{sec:discrete.open})
\begin{equation}
   \gamma_{\rm class} = \frac{\pi^2}{2 L^2}.
\end{equation}
It is interesting to compare this result with those obtained by Barra
and Gaspard~\cite{barra99da} in their study of scattering resonances
for an open, periodic chain of scatterers. In high energy limit they
found that the logarithms of the magnitudes of the eigenvalues can be
bounded above and below by functions that scale as $1/L$.  They expect
that the lower bound, given by the eigenvalues in the middle of the
band, should hold also for lower energies~\footnote{P.  Gaspard,
  private communication}. On the other hand, the upper bound, which
gives the escape rate, is given by the resonances near the edges of
the bands which are harder to estimate at low energies.  Therefore
this bound is more difficult to control.

This reasoning is consistent with our findings. In our case, the
eigenvalues of the smallest magnitude are those for which to $k\approx
\pm L/2$ (the middle of the band; see Figure~\ref{fig:abv}). Thus
their magnitude can be estimated from~(\ref{eq:approx_v}) setting
$a=0, b=1$ and therefore their logarithms scale as $1/L$. On the other
hand, the eigenvalues of largest magnitude, which give the escape
rate, lie at the edges of the band.

The discrepancy between our results is not surprising for the
high-energy limit corresponds to semi-classical limit for our system,
and in the present work we consider the extreme quantum case.

\subsection{Steady state solution}
\label{sec:steady_reg}

Suppose now that the multibaker of length $L$ is connected at both
ends to infinitely conducting leads. We suppose that there can be
traveling waves in the leads moving to the right and to the left.
These waves are most conveniently described in terms of the momentum
space representation of the wave functions, and we recall that the
``bottom'' states come from the left and the ``top'' states come from
the right.
\begin{figure*}[ht]
  \centering
  \includegraphics[scale=0.5]{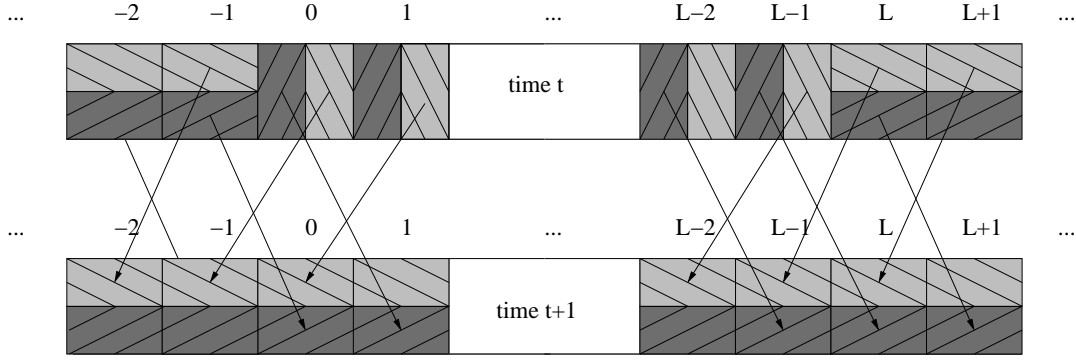}
  \caption{Scattering from a quantum multibaker.}
  \label{fig:scatterqmb}
\end{figure*}
Thus to the left of
the chain, $n<0$, we take the traveling waves to be
\begin{eqnarray}
  \widetilde\Psi_b(n,t) & = & A e^{i(\omega t - kn)}, \\
  \widetilde\Psi_t(n,t) & = & B e^{i(\omega t + kn)},
\end{eqnarray}
and to the right of the chain, $n>L-1$, the waves are
\begin{eqnarray}
  \widetilde\Psi_b(n,t) & = & C e^{i(\omega t - kn)}, \\
  \widetilde\Psi_t(n,t) & = & D e^{i(\omega t + kn)}.
\end{eqnarray}
Here $A,D$ are the amplitudes of the incoming waves, while $B,C$ are
the amplitudes of the outgoing waves. Due to the dynamics on the
multibaker, we can match the incoming wave functions in the leads to
the proper momentum space functions for the unit cells at $0$ and at
$L-1$. This matching condition is simply
\begin{eqnarray}
  \widetilde\Psi_b(0,t) & = & A e^{i\omega t }, \label{eq:65}\\
  \widetilde\Psi_t(L-1,t) & = & D e^{i\omega t}e^{ i(L-1)k}.
\end{eqnarray}
We will use a scattering approach to find the outgoing amplitudes,
$B,C$ for the steady state solution, as well as to solve the problem
of the relaxation of some initial state to a steady state. First we
consider the steady state solution for the baker chain with conducting
leads.

The steady state solution is defined by the condition that the time
dependence of the wave function can be incorporated in a time
dependent phase factor. Since the dynamics takes place at discrete
times, there is a $\widetilde\omega$ such that $\Psi_{t,b}(n,t+1) =
e^{i\widetilde\omega} \Psi_{t,b}(n,t)$ which implies that
$\Psi_{t,b}(n,t) = e^{i\widetilde\omega t} \Psi_{t,b}(n,0)$.  In
particular, using Eq.~(\ref{eq:65}),
$\Psi_{b}(0,t)=e^{i\widetilde\omega t} \Psi_{b}(0,0)$, so that
$\widetilde\omega\equiv\omega$.  Writing $\widetilde\Psi_b(n,t) =
e^{i\omega t}\widetilde\Psi_b(n)$, etc., we obtain the steady state
equation
\begin{equation}
  \label{eq:propomega}
  \left[
    \begin{array}[c]{c}
      \widetilde\Psi_b(n+1) \\ \widetilde\Psi_t(n-1)
    \end{array}
  \right] =
  \left[
    \begin{array}[c]{cc}
      f_0 e^{-i\omega} & 0 \\ 
     0 & f_0 e^{-i\omega}
    \end{array}
  \right]
  \left[
    \begin{array}[c]{c}
      \Psi_l(n) \\ \Psi_r(n)
    \end{array}
  \right] .
\end{equation}

The transmission and reflection coefficients for the chain
can be expressed in terms of the scattering $S$-matrix, given by
\begin{equation}
   \left[
    \begin{array}[c]{c}
      \widetilde\Psi_t(-1) \\ \widetilde\Psi_b(L)
    \end{array}
  \right] =
  S_{0,L-1}
  \left[
    \begin{array}[c]{c}
      \widetilde\Psi_b(0) \\ \widetilde\Psi_t(L-1)
    \end{array}
  \right] ,
\end{equation}
where the elements of $S$-matrix are
\begin{equation}
  S_{0,L-1} = \left[
    \begin{array}[c]{cc}
      r_{0,L-1} & t'_{0,L-1} \\ 
      t_{0,L-1} & r'_{0,L-1}
    \end{array}
  \right].
\end{equation}
Here the unprimed coefficients refer to waves incident on the left end
of the chain, while the primed quantities refer to the waves incident
on the right side of the chain.  The transmission and reflection
coefficients, $T, T', R, R'$ respectively, are then obtained from the
elements of $S$ by
\begin{equation}
  T=|t_{0,L-1}|^2, \qquad R=|r_{0,L-1}|^2,
\end{equation}
and similarly for the primed quantities. Unitarity of $S$ implies
$T=T'$, $R=R'$.  In order to calculate the $S$-matrix, $S_{0,L-1}$,
for the chain, we proceed as for the absorbing case, by looking at the
transfer and scattering matrices for one cell, and building up the
matrices for the chain by iteration, cell by cell.  Consider the cell
labelled by the index $n$. The $S$-matrix for the $n$-th cell is given
by
\begin{equation}
  \label{eq:smatrixn2}
   \left[
    \begin{array}[c]{c}
      \widetilde\Psi_t(n-1) \\ \widetilde\Psi_b(n+1)
    \end{array}
  \right] =
  S_n
  \left[
    \begin{array}[c]{c}
      \widetilde\Psi_b(n) \\ \widetilde\Psi_t(n)
    \end{array}
  \right] ,
\end{equation}
and the transfer $T$-matrix is 
\begin{equation}
   \left[
    \begin{array}[c]{c}
      \widetilde\Psi_b(n+1) \\ \widetilde\Psi_t(n)
    \end{array}
  \right] =
  T_n
  \left[
    \begin{array}[c]{c}
      \widetilde\Psi_b(n) \\ \widetilde\Psi_t(n-1)
    \end{array}
  \right] .
\end{equation}
Each of the matrices can be given in terms of the other, thus,
\begin{eqnarray}
  S = \left[
    \begin{array}[c]{cc}
      r & t' \\ 
      t & r'
    \end{array}
  \right]
  & \Rightarrow &
  T = \left[
    \begin{array}[c]{cc}
      t-r' {t'}^{-1} r & r' {t'}^{-1} \\
      -{t'}^{-1} r & {t'}^{-1} 
    \end{array}
  \right],\\
  T = \left[
    \begin{array}[c]{cc}
      \alpha & \gamma \\
      \beta & \delta
    \end{array}
  \right]
  & \Rightarrow &
  S = \left[
    \begin{array}[c]{cc}
      -\delta^{-1} \beta & \delta^{-1} \\
      \alpha-\gamma\delta^{-1}\beta & \gamma\delta^{-1}
    \end{array}
  \right].
\end{eqnarray}

The $S$ and $T$ matrices can easily be obtained by transforming the
dynamical equations~(\ref{eq:propomega}) to momentum
representation~(\ref{eq:b9}), so that
\begin{equation}
    \left[
    \begin{array}[c]{c}
      \widetilde\Psi_b(n+1) \\ \widetilde\Psi_t(n-1)
    \end{array}
  \right] = f_0 e^{-i\omega}
  \left[
    \begin{array}[c]{cc}
      g_{00} & g_{01}  \\
      g_{10}  & g_{11}
    \end{array}
  \right]
  \left[
    \begin{array}[c]{c}
      \widetilde\Psi_b(n) \\ \widetilde\Psi_t(n)
    \end{array}
  \right] ,
\end{equation}
from which the $S$-matrix follows as
\begin{equation}
  S_n =
  \frac{e^{-i\omega}}{\sqrt{2}}
  \left[
    \begin{array}[c]{cc}
      e^{i\pi\phi_p(1-\phi_q)} &       
      e^{i\pi(1+\phi_q+\phi_p-\phi_q\phi_p)} \\
      e^{-i\pi\phi_q \phi_p} &      
      e^{i\pi\phi_q(1-\phi_p)}
    \end{array}
  \right].
\end{equation}
The $T$-matrix is then given by
\begin{equation}
  T_n =
  \left[
    \begin{array}[c]{cc}
      \sqrt{2} e^{-i\omega}e^{-i\pi\phi_q \phi_p} & 
      -e^{-i\pi \phi_p} \\
      e^{-i \pi \phi_q} & 
      \sqrt{2} e^{i\omega} e^{-i\pi(1+\phi_q+\phi_p-\phi_q \phi_p)}
    \end{array}
  \right].
\end{equation}
The scattering matrix for the whole multibaker $S_{0,L-1}$ can easily
be derived from $T_{0,L-1}:=T_{L-1}\cdot \dots \cdot T_1 \cdot T_0$.
Its unitarity can also be verified. For the uniform system
\begin{equation}
  \label{eq:tmatrixL}
  T_{0,L-1}=T^L = \frac{\chi_+^L-\chi_-^L}{\chi_+-\chi_-}T
  -\frac{\chi_-\chi_+^L-\chi_+\chi_-^L}{\chi_+-\chi_-}
\end{equation}
where $\chi_\pm$ are roots of characteristic polynomial of $T$
\begin{equation}
  \chi_\pm = e^{-i\alpha}[\sqrt{2}\cos(\beta-\omega)\pm\sqrt{\cos
  2(\beta-\omega)}],
\end{equation}
and $\alpha,\beta$ are given by Eq.~(\ref{eq:alphabeta}).
Depending on the sign of $\cos 2(\beta+\omega)$ there are two types of
solutions: if the frequency of the incident wave falls in one of the
quasi-energy bands
\begin{equation}
  \cos 2(\beta-\omega) <0 \Leftrightarrow
  \omega-\beta\in[\pi/4,3\pi/4]\cup[5\pi/4,7\pi/4], 
\end{equation} 
we have the oscillatory case with some interesting structure.
Otherwise, when the frequency of the incident wave falls in the gap,
we observe almost total reflection of particles coming from the leads
to the chain, becoming total as $L\rightarrow\infty$ (the exponential
case).
\begin{enumerate}
\item If $\cos 2(\beta-\omega)<0$ (oscillatory case), the
  characteristic roots are: \\
  \begin{equation}
    \chi_\pm = e^{-i\alpha}[\sqrt{2}\cos(\beta-\omega)\pm
    i\sqrt{-\cos 2(\beta-\omega)}],
  \end{equation}
  thus $|\chi_\pm|^2 =1$.  Set \( \chi_\pm = e^{-i\alpha}e^{\pm
    i\kappa}.\) Then the scattering matrix for the chain becomes,
  \begin{equation}
    S_{0,L-1}=
    \frac{1}{z_L}
    \left[
      \begin{array}[c]{cc}
        -\sin L\kappa\, e^{i(\alpha- \pi \phi_q)} & 
        \sin \kappa\, e^{i\alpha L}\\
        \sin \kappa\, e^{-i\alpha L}& 
        -\sin L\kappa \, e^{i(\alpha-\pi \phi_p)}
      \end{array}
    \right].
    \label{eq:77}
  \end{equation}
  To simplify the formulas we introduce
  \begin{eqnarray*}
    z_n \equiv r_n e^{i \phi_n} & := & \sqrt{2} \sin n\kappa e^{-i(\beta-\omega)} - \sin
    \kappa(n-1) \\
    & = & \cos \kappa n \sin \kappa 
    -i\epsilon \sin n\kappa \sqrt{1+\sin^2 \kappa},
  \end{eqnarray*}
  where $\epsilon=\pm$ is the sign of $\sin (\beta-\omega)$. 
  Then the transmission and reflection coefficients are
  \begin{eqnarray}
    R 
    & = & \frac{\sin^2 L\kappa}{\sin^2 \kappa +
      \sin^2 L\kappa} = \frac{1}{1+\frac{\sin^2 \kappa}{\sin^2 L\kappa}},
    \\
    T 
    & = & \frac{\sin^2 \kappa}{\sin^2 \kappa +
      \sin^2 L\kappa} = \frac{1}{1+\frac{\sin^2 L\kappa}{\sin^2 \kappa}}.
  \end{eqnarray}
  Some interesting special cases occur when:
  \begin{enumerate}
  \item $\kappa=k\pi + \pi/2$, $L$ odd: $T=1/2$; 
  \item $\kappa=k\pi + \pi/2$, $L$ even: $T=1$ ;
  \item $\kappa=k\pi$: $T=1/(1+L^2)$ ;
  \item $\kappa=k\pi/L$: $T=1$.
  \end{enumerate}
  We will refer to the cases when $T=1$ as transmission resonances.
  They occur when $\sin L\kappa = 0$. On the other hand, one can see
  from Eq.~(\ref{eq:77}) that the S-matrix has poles when
  Eq.~(\ref{eq:sin2}) is satisfied. Hence the poles of the S-matrix
  determine the eigenstates of open system.

\item In the exponential case, when $\cos 2(\beta-\omega)>0$, we have
  $|\chi_\pm|^2\geq 1 $, and $\chi_+ \chi^\ast_- = 1$, so that \(
  |\chi_-| = \frac{1}{|\chi_+|}.\) Then the transmission and
  reflection coefficients are
  \begin{eqnarray}
    R & = & \frac{(|\chi_+|^L -
      |\chi_-|^L)^2}{(|\chi_+|^L - |\chi_-|^L)^2
      +(|\chi_+| - |\chi_-|)^2} \\
    & \approx & 1-|\chi_-|^{2(L-1)}
    \approx 1\\
    T & = & \frac{(|\chi_+| -
      |\chi_-|)^2}{(|\chi_+|^L - |\chi_-|^L)^2+
      (|\chi_+| - |\chi_-|)^2} \\
    &\approx& |\chi_-|^{2(L-1)} 
    \approx 0.
  \end{eqnarray}
\end{enumerate}

\subsection{Density profile in the steady state --- violation of the
  Fick's law}

As mentioned above, the oscillatory case provides some interesting
structures, illustrating the interference between waves traveling to
the right and left along the chain. The algebra is tedious but
straightforward, and we don't reproduce it here, merely stating the
final results.

\begin{figure}[ht]
  \centering
  \begin{tabular}[c]{c}
  \includegraphics[scale=0.5]{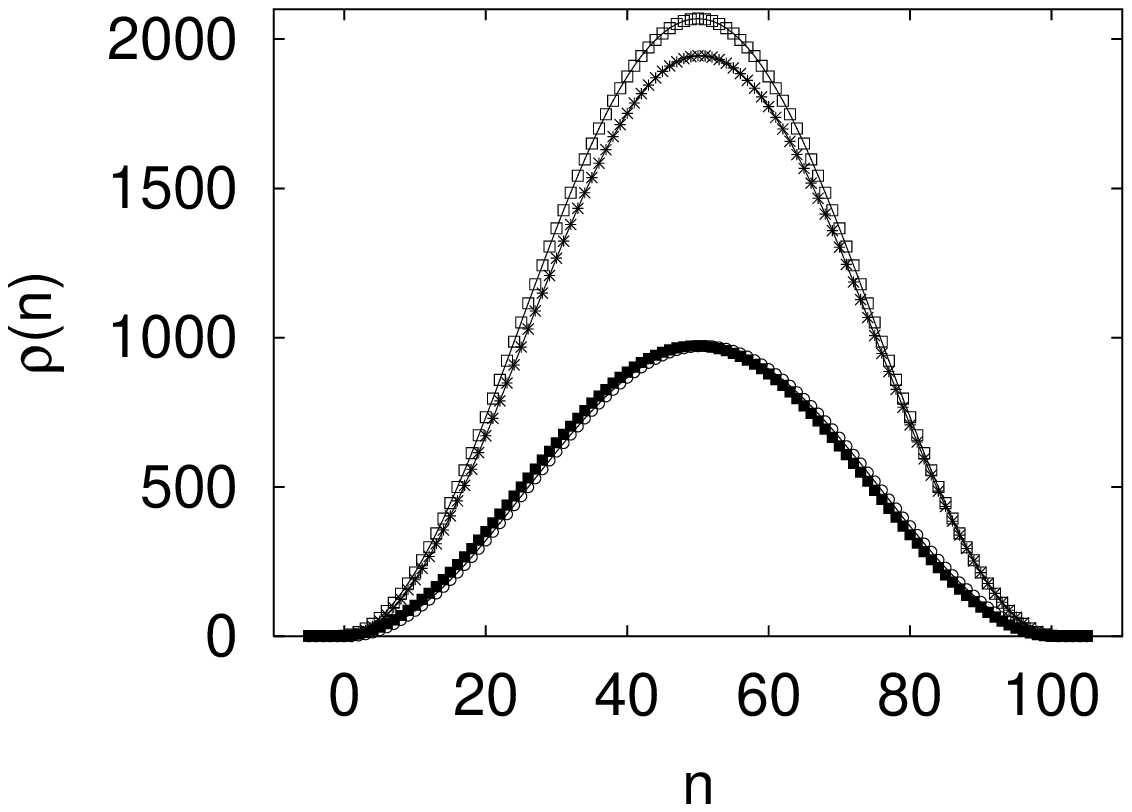} \\
  \includegraphics[scale=0.5]{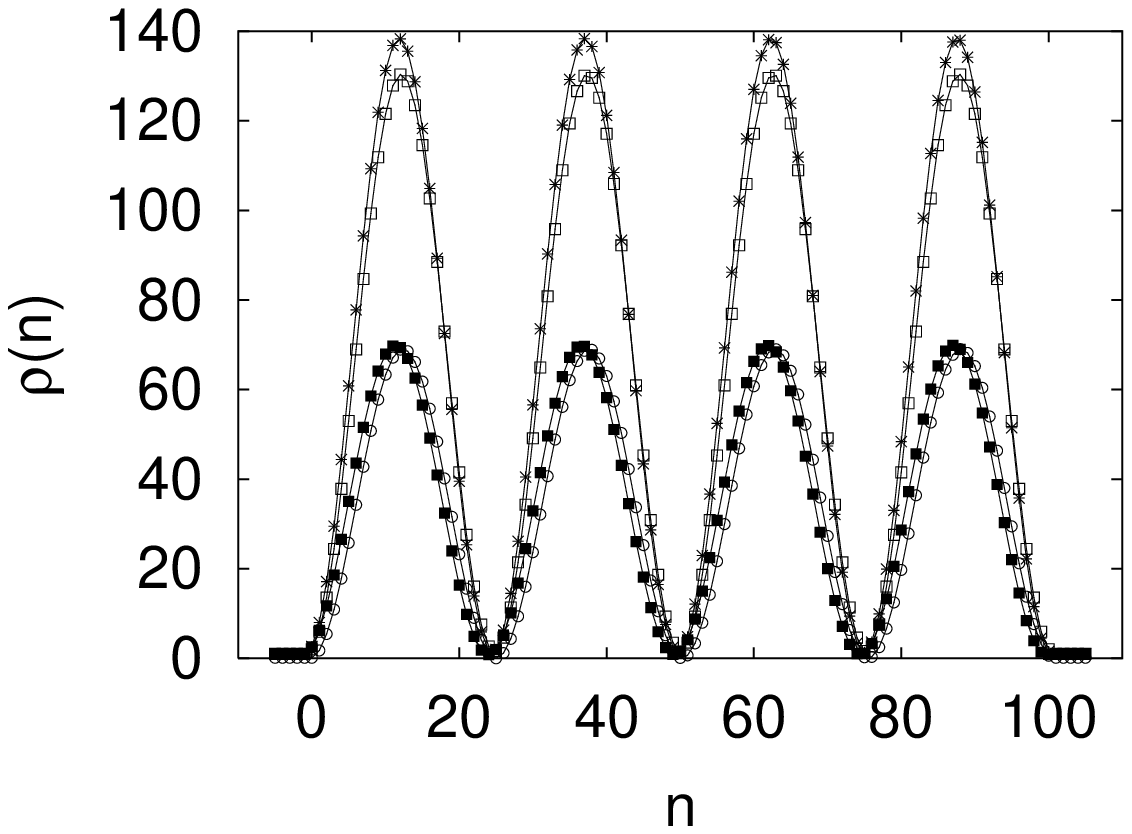}
  \end{tabular}
  \caption{Profile of the probability density in the steady state for
    the smallest and the fourth resonances $\kappa=k \pi/L$ with
    $k=1,4$.  Open boxes show the approximate solution, stars stand
    for the full solution. Also shown is the probability density of
    the bottom (full boxes) and the top states (open circles). We took
    $\rho_L=0.1, \rho_R=0.9, \eta=0$. The horizontal axis range is
    $[-5,105]$.}
  \label{fig:steady_res}
\end{figure}

The wave function in the steady state is
\begin{widetext}
\begin{eqnarray*}
  \Psi_b(n) & = & \frac{e^{-i\alpha n}}{z_L}
  [z_{L-n}\Psi_b(0) -e^{i\alpha L} e^{i(\alpha-\pi \phi_p)} \sin
  n\kappa\, \Psi_t(L-1)], \\
  \Psi_t(n) & = & \frac{e^{-i\alpha (n+1-L)}}{z_L}
  [e^{-i\alpha L}e^{i(\alpha- \pi \phi_q)} \sin
  \kappa(n+1-L)\,\Psi_b(0)+z_{n+1} \Psi_t(L-1) ].
\end{eqnarray*}
\end{widetext}
We introduce the probability densities, $\rho_L$ and $\rho_R$ from the
left and right leads, respectively, in terms of the corresponding wave
functions, by writing \( \Psi_b(0) = \sqrt{\rho_L}, \, \Psi_t(L-1) =
\sqrt{\rho_R} e^{i\eta},\) where $\eta$ denotes a relative phase
between the wave functions at the two ends. Then,
introducing the angle \( \phi = \pi(\phi_q - \phi_p)/2 +\alpha L
+\eta,\) we obtain the total probability density at cell $n$
\begin{widetext}
\begin{eqnarray*}
  \rho(n) & = & \frac{\sin^2 (L-n-1)\kappa+\sin^2 (L-n)\kappa+\sin^2
  \kappa}{|z_L|^2} \rho_L + \frac{\sin^2 \kappa n+\sin^2 \kappa (n+1)+\sin^2
  \kappa}{|z_L|^2} \rho_R\\ 
  & & - i\frac{ \sqrt{\rho_L \rho_R}}{|z_L|^2} \left\{
  \sin (L-1-n)\kappa\,[z_{n+1} e^{i\phi} -z_{n+1}^\ast e^{-i\phi}
  ]  - \sin n\kappa \,
  [z_{L-n}^\ast e^{i\phi} -z_{L-n} e^{-i\phi}]\right\}. 
\end{eqnarray*}
\end{widetext}
At resonance ($\kappa=k \pi/L$) it takes form
\begin{eqnarray}
  \rho(n) &= &\left( 1+\frac{\sin^2 \kappa n+\sin^2 \kappa (n+1)}{\sin^2
      \kappa}\right) (\rho_L+\rho_R) \nonumber\\
  &&+ 2\frac{ \sqrt{\rho_L \rho_R}}{\sin \kappa}
  r_{2n+1} \sin(\phi+\phi_{2n+1}). \label{eq:rho_res} 
\end{eqnarray}
Let us concentrate on this last expression, for simplicity.  Since \(
|z_n|^2 = \sin^2 \kappa + \sin^2 n \kappa,\) then if we write \(
\rho(n) = \rho_1(n) (\rho_L+\rho_R) + 2 \rho_2(n) \sqrt{\rho_L
  \rho_R}, \) then \( 0 \leq \frac{|\rho_2|}{\rho_1}\frac{2
  \sqrt{\rho_L \rho_R}}{\rho_L+\rho_R} \leq 1 ,\) which implies, in
particular, positivity of $\rho$. For small $k$ it turns out that the
second term is negligible~(Figure~\ref{fig:steady_res}).
We can easily estimate the behavior of the probability density profile
in this case. For large $L$ and small $k$ we have
\begin{equation}
  \label{eq:rho_approx1}
  \rho_k(n) \approx 
  \left( 1+\frac{1-\cos k\pi/L}{k^2\pi^2/L^2}
    \cos\frac{k\pi(2n+1)}{L}
  \right) (\rho_L+\rho_R),
\end{equation}
which for $k=1$ can also be approximated as \( \rho(n) \approx \left(
  1+\frac{2L^2}{\pi^2} \sin\frac{\pi}{L}(n+\frac12) \right)
(\rho_L+\rho_R).  \) Figure~\ref{fig:steady_res} shows the approximate
solution (crosses) and the full solution (diamonds) as well as the
probability density of the bottom states (boxes) and the top states
(circles).  For the smallest resonance ($k=1$) the probability
distribution achieves maximum around $n=L/2$ where it is approximately
$2 L^2/\pi^2$.

These results are clearly connected to the slow probability escape
$\propto 1/L^3$. To understand them consider a plane wave coming from
the left with a resonant frequency going through the open quantum
multibaker. Thus at every time step we inject the same density inside.
The wave travels ballistically inside and when it reaches the end is
mostly reflected, partially transmitted. Due to the fast motion inside
and slow decay the density accumulates in the multibaker and reaches
the steady state when the escape on the right balanced the injection
on the left. The probability density of the resulting standing wave is
given by Eq.~(\ref{eq:rho_res}).

This result is very striking in comparison with the classical case: in
the classical multibaker one obtains Fick's
behavior~\cite{gaspard92s,tasaki95s,Gaspard98sa} --- there is a linear
profile of probability density. This is also what happens for
partially integrated classical dynamics which we considered in
section~\ref{sec:discrete.reservoirs}. In particular, the probability
density at any point inside the multibaker is between the densities of
the reservoirs.

We defer the complete discussion of the steady state solutions to a
further work where it will be considered together with the
semi-classical case in the context of transport~\cite{wojcik02de}.
Here let us only mention that the approach to the steady state can be
conveniently studied as a spectral problem: The evolution equations
for the quantum multibaker with two waves scattering from left and
right can be written as
\begin{equation}
  \widehat{\Psi}(t) = \widehat{\mathbf M}_L \widehat{\Psi}(t-1) + \Phi_0,
\end{equation}
where $ \widehat{\Psi}(t) := e^{-i\omega t} \Psi(t)$, $\widehat{\mathbf M}_L$
is the matrix representation of the open multibaker propagator
following from the equations
\begin{equation}
  \left[
    \begin{array}[c]{c}
      \widehat \Psi_l(n,t) \\ \widehat\Psi_r(n,t)
    \end{array}
  \right] = 
  G_2^{-1} 
  \cdot
  \left[
    \begin{array}[c]{cc}
      f_0 e^{-i\omega}& 0\\
      0& f_0 e^{-i \omega}
    \end{array}
  \right]
  \left[
    \begin{array}[c]{c}
      \widehat\Psi_l(n-1,t) \\ \widehat\Psi_r(n+1,t)
    \end{array}
  \right]  ,
\end{equation}
and $\Phi_0$ denotes the steady state boundary conditions: $\Phi_0 =
[\Phi_b(0), \Phi_t(0), \dots, \Phi_b(L-1), \Phi_t(L-1)]^T$, $\Phi_b(0)
= A$, $\Phi_t(L-1) = De^{ik(L-1)}$, $\Phi_{b,t}(n) = 0 \;\; {\rm
  otherwise}$.  The solution to this simple affine problem is
\begin{equation}
  \label{eq:steady_approach}
  |\widehat\Psi(t)\rangle = \sum_{\lambda_k} \frac{1-\lambda_k^t}{1-\lambda_k}
  |\phi_k\rangle\langle \phi_k| \Phi_0\rangle   
  + \sum_{\lambda_k} \lambda_k^t  |\phi_k\rangle\langle \phi_k| \widehat\Psi(0)\rangle,
\end{equation}
where $\lambda_k$ are the $\omega$-dependent eigenvalues of
$\widehat{\mathbf M}_L$ and the $|\phi_k\rangle$ are the corresponding
eigenvectors.  In particular, if at time 0 the system is empty
$\Psi(0)=0$, then the solution is
\begin{equation}
  \label{eq:steady_approach2}
  |\widehat\Psi(t)\rangle = \sum_{\lambda_k} \frac{1-\lambda_k^t}{1-\lambda_k}
  |\phi_k\rangle\langle \phi_k| \Phi_0\rangle  . 
\end{equation}
The steady state is the time invariant part of the above solution
\begin{equation}
  |\widehat\Psi\rangle = \sum_{\lambda_k} \frac{1}{1-\lambda_k}
  |\phi_k\rangle\langle \phi_k| \Phi_0\rangle.
\end{equation}
The approach to the steady state is given by the eigenvalues of the
open multibaker~(\ref{eq:approx_v}), thus it is as slow as the escape
of probability density, which is consistent with the accumulation of
large probability density in the system.  Note that the distribution
of the absolute values of the eigenvalues of $\widehat{\mathbf M}_L$ is
$\omega$ independent, yet the steady state solution does depend on
$\omega$.

\section{The closed random quantum multibaker for $N=2$}
\label{sec:random}
In this section we extend our discussion of the quantum multibaker map
for $N=2$ by considering the case where the phases, $\phi_{q,p}$,
defining the map vary randomly from cell to cell. As expected, the
random case differs considerably from the uniform case, since the
randomness of the phases acts as a disordering mechanism producing a
localization of the wave function. Unlike the uniform case, there is
little that can be done analytically for the random case, other than
making use of some known results for the properties of products of
random $2\times2$ matrices \cite{crisanti93s}, which in this case are
only of limited utility. For this reason we limit ourselves to a
presentation of the results of numerical studies.

The random quantum multibaker map is, for the case $N=2$, is defined by
the equations
\begin{eqnarray*}
  \Psi_l(n,t+1) & = & g_{00}(n) f_0(n-1) \Psi_l(n-1,t) \\
  &&+  g_{01}(n)
  f_0(n+1) \Psi_r(n+1,t), \\ 
  \Psi_r(n,t+1) & = & g_{10}(n) f_0(n-1)\Psi_l(n-1,t) \\
  &&+  g_{11}(n)
  f_0(n+1) \Psi_r(n+1,t),  
\end{eqnarray*}
where the phases in each of the cells are drawn randomly from some
distribution. Here we use a uniform distribution of phases in the unit
interval.

\begin{figure}[htbp]
  \centering
  \includegraphics[scale=0.5]{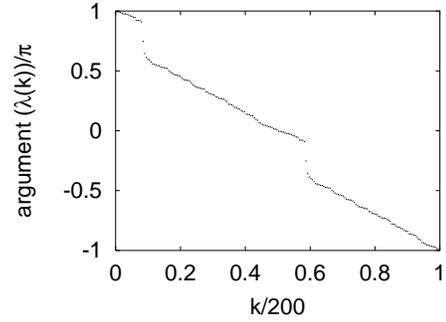}
  \caption{Spectrum of a the quantum multibaker map for $L=101$ cells.} 
  \label{fig:random_spectrum}
\end{figure}

The numerically obtained quasi-energy spectrum is illustrated
in~Figure~\ref{fig:random_spectrum} and can be compared with that for
the uniform case.
The quasi-energies associated with the eigenstates \( \Psi(t) =
[\Psi_l(0), \Psi_r(0), \dots, \Psi_l(L-1), \Psi_r(L-1)], \) are
determined by the solution of the following eigenvalue equation,
\begin{equation}
  \lambda
  \left[
    \begin{array}[c]{c}
      \Psi_{l}(n)\\ \Psi_{r}(n)
    \end{array}    
  \right] =  
  \left[
    \begin{array}[c]{cc}
      g_{00}(n) & g_{01}(n)\\
      g_{10}(n) & g_{11}(n)
    \end{array}
  \right] 
  \left[
    \begin{array}[c]{c}
      f_0(n-1)\Psi_{l}(n-1)\\ f_0(n+1)\Psi_{r}(n+1)
    \end{array}    
  \right],
\end{equation}
which determines the eigenvalue $\lambda$.  It is interesting to note
that this equation can be put into a form which is reminiscent of the
Anderson model for localization (see appendix~\ref{app:anderson}).  If
we define $\widehat\Psi(k)$ by \(
\widehat{\Psi}_{l}(k):=f_0(k){\Psi}_{l}(k) , \,
\widehat{\Psi}_{r}(k):=f_0(k){\Psi}_{r}(k),\) we can obtain a set of
equations that define a generalized Anderson model:
\begin{widetext}
\begin{equation}
  \left[\frac{\lambda  g^*_{10}(n+1)}{f_0(n)} - \frac{f_0(n+1)
      g_{01}(n)}{\lambda} \right]
  \widehat{\Psi}_{l}(n) = g_{00}(n)g^*_{10}(n+1)\widehat{\Psi}_{l}(n-1) -
  g^*_{00}(n+1)g_{01}(n) \widehat{\Psi}_{l}(n+1) . 
\end{equation}
\end{widetext}
A similar equation holds for $\widehat{\Psi}_r(n+1)$.  We rewrite this
equation so that it takes the form a dynamical problem, where the cell
index $n$ plays the role of the time step. That is, \(
\widehat{\Psi}_{l}(n+1) = 2\sqrt{2} e^{i\phi_1} \sin(\phi_3)
\widehat{\Psi}_{l}(n) - e^{i\phi_2}\widehat{\Psi}_{l}(n-1) , \) or,
using transfer matrices,
\begin{equation}
  \label{eq:fibonnacci}
    \left[
    \begin{array}[c]{c}
      \widehat{\Psi}_{l}(n+1)\\ \widehat{\Psi}_{l}(n)
    \end{array}    
  \right] = 
  \left[
    \begin{array}[c]{cc}
      2\sqrt{2} e^{i\phi_1}
      \sin(\phi_3) & -e^{i\phi_2}\\
      1  & 0 
    \end{array}
  \right] 
  \left[
    \begin{array}[c]{c}
      \widehat{\Psi}_{l}(n)\\ \widehat{\Psi}_{l}(n-1)
    \end{array}    
  \right],
\end{equation}
where the transfer matrix can be written as 
\begin{equation}
\label{eq:transf_decomp}
  \left[
    \begin{array}[c]{cc}
      e^{i\phi_1} & 0\\
      0 & 1 
    \end{array}
  \right] 
  \left[
    \begin{array}[c]{cc}
      2\sqrt{2} \sin(\phi_3) & -1\\
      1  & 0 
    \end{array}
  \right] 
  \left[
    \begin{array}[c]{cc}
      1 & 0\\
      0 & e^{i\phi_2}
    \end{array}
  \right] .
\end{equation}
The phases are given by
\begin{eqnarray*}
  \phi_1 & = & (\pi/2)[-1+\phi_q(n)\phi_p(n) - \phi_q(n+1)\phi_p(n+1)\\
  && -\phi_q(n) -\phi_p(n+1)], \\
  \phi_2 & = & -\pi(\phi_q(n)+\phi_p(n+1)+1) ,\\
  \phi_3 & = & \kappa + (\pi/2)[\phi_q(n)\phi_p(n) + \phi_q(n+1)\phi_p(n+1)\\
  &&-\phi_q(n) -\phi_p(n+1)].
\end{eqnarray*}

\begin{figure}[htbp]
  \centering
  \begin{tabular}[c]{cc}
  \includegraphics[scale=0.35]{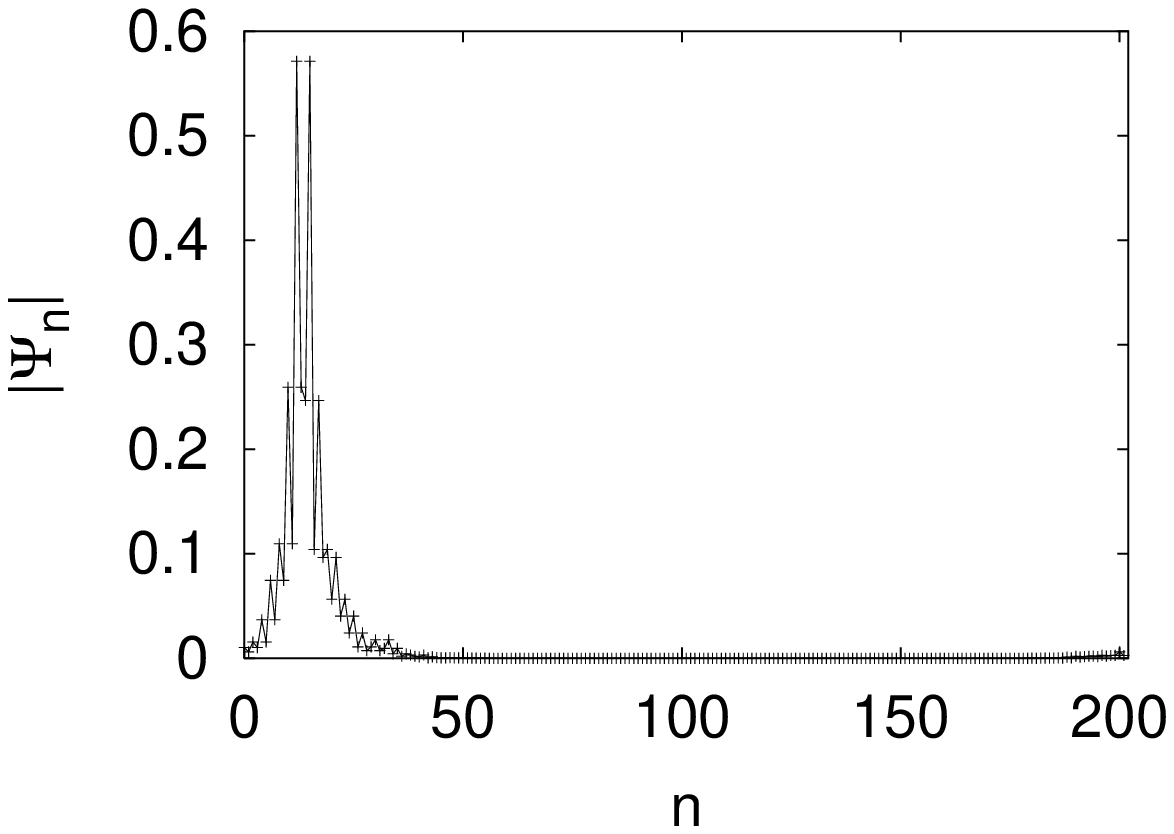} &
  \includegraphics[scale=0.35]{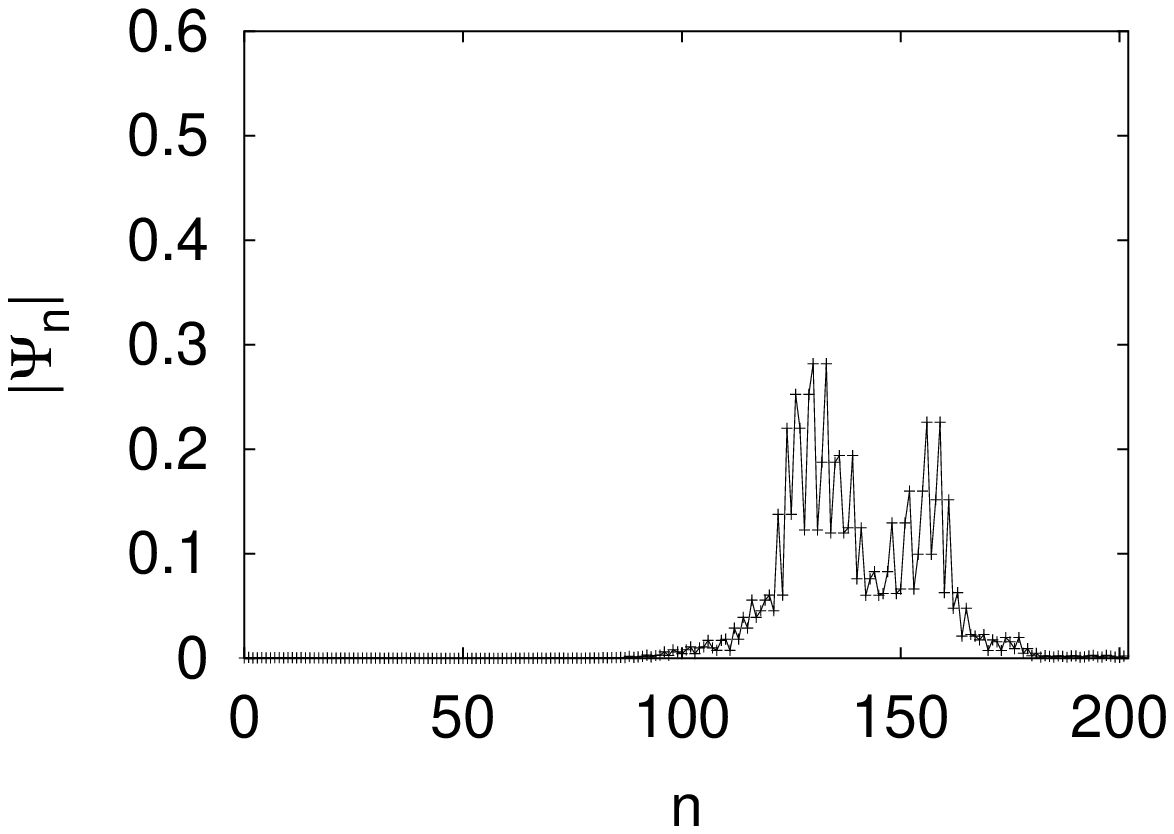} \\
  \includegraphics[scale=0.35]{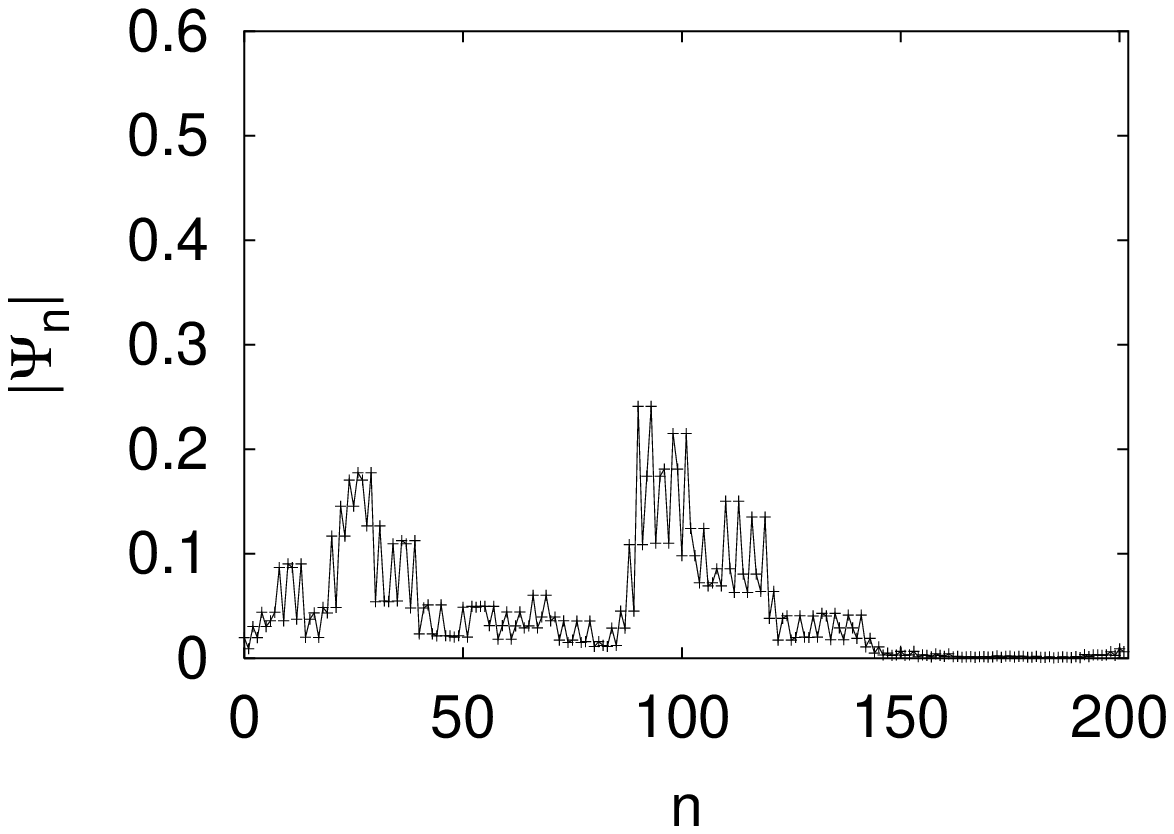} &
  \includegraphics[scale=0.35]{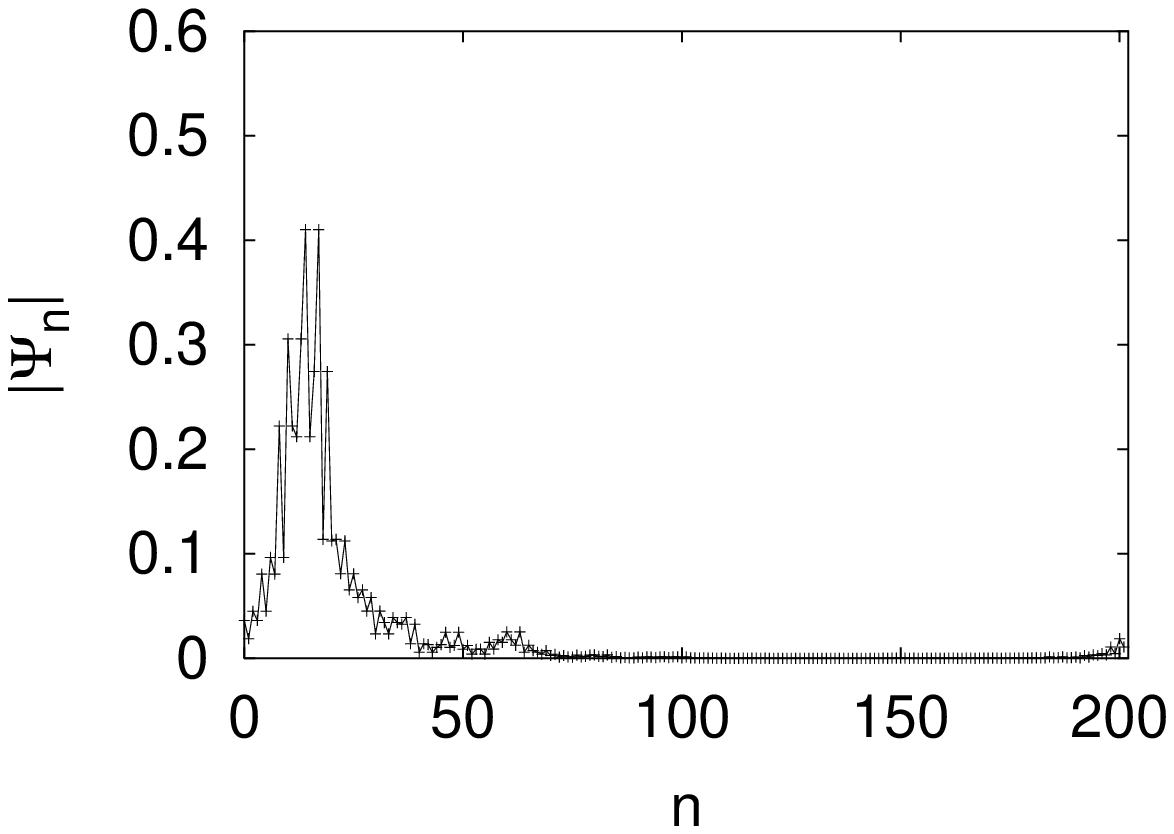}
  \end{tabular}
  \caption{Four examples out of 202 eigenstates of a realization of
    the quantum random multibaker of length $L=101$ with periodic
    boundary conditions. Absolute values of $\Psi_l(n), \Psi_r(n)$ are
    shown. Notice that $|\Psi_l(n)|=|\Psi_r(n+1)|$ which results in
    particular in the twin peak structures discussed in text.}
  \label{fig:random_states}
\end{figure}

\begin{figure}[htbp]
  \centering
  \includegraphics[scale=0.5]{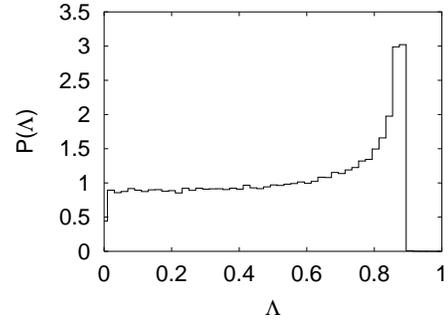}
  \caption{Distribution of the ratios  $\Lambda := \left| \ln
      \frac{|\widehat{\Psi}_{l}(n+1)|}{|\widehat{\Psi}_{l}(n)|}
    \right|$ for a set of 2402 eigenstates of a realisation of a
    quantum random multibaker of length 1201. }
  \label{fig:liap1}
\end{figure}

It can be seen that the eigenstates are localized but the localization
does not seem to be purely exponential for finite $L$, as illustrated
in~Figure~\ref{fig:random_states}, where some states are localized
over some tens of cells, while others are localized over several times
as many cells. In longer chains the exponential decay of the
eigenfunctions is more pronounced.
The exponential decay of the wavefunction far from its peak is usually
characterized by the inverse localization length
\begin{equation}
  \xi_0^{-1} := \lim_{|n|\rightarrow \infty} \ln
  \frac{|\Psi_n|}{|\Psi_{n+1}|} .
\end{equation}
A simple estimate can be obtained as follows: From
Eq.~(\ref{eq:fibonnacci}) we get \( |\widehat{\Psi}_{l}(n+1)| =
|2\sqrt{2} e^{i(\phi_1-\phi_2)} \sin(\phi_3) \widehat{\Psi}_{l}(n) +
\widehat{\Psi}_{l}(n-1)| .  \) Thus, on the average we have \(
\langle|\widehat{\Psi}_{l}(n+1)|^2\rangle = 4
\langle|\widehat{\Psi}_{l}(n)|^2\rangle +
\langle|\widehat{\Psi}_{l}(n-1)|^2\rangle.  \) Therefore, starting
from almost every initial conditions, on the average we should observe
growth of $|\widehat{\Psi}_{l}(n)|^2$ given by \(
|\widehat{\Psi}_{l}(n+1)|/|\widehat{\Psi}_{l}(n)| \approx
\sqrt{2+\sqrt{5}}\approx 2.06 .  \) Therefore the inverse localization
length is approximately 0.72. Of course, rather than calculate the
logarithm of the average we should calculate the average of the
logarithm but the obtained value is not far off the numerically
obtained average, which is 0.51 for a chain of 301 cells, and 0.56 for
a chain of 1201 cells.  Figure~\ref{fig:liap1} shows the distribution
of the numbers $|\ln \frac{|\Psi_n|}{|\Psi_{n+1}|}|$ over the range of
the wavefunction where it was appreciably different from 0, over all
of the eigenstates for a given realization of disorder.

This distribution reflects the generally broad distributions
associated with the properties of localized states. The difference
between the estimated value of the rate of growth with the average
obtained from the numerical distribution is due to the contributions
from regions where the variation in amplitude from cell to cell is not
exponential (compare with Fig.~\ref{fig:random_states}).

\begin{figure}[htbp]
  \centering
  \begin{tabular}[c]{cc}
    \includegraphics[scale=0.35]{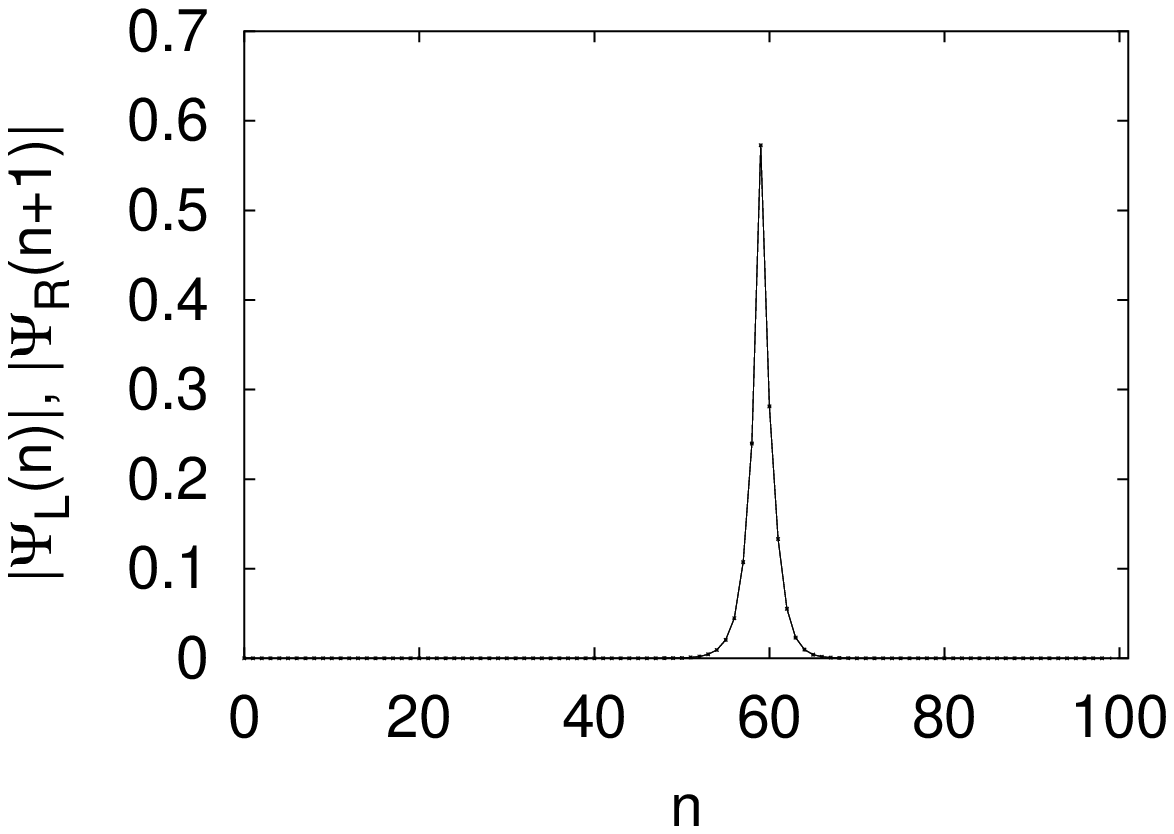} &
    \includegraphics[scale=0.35]{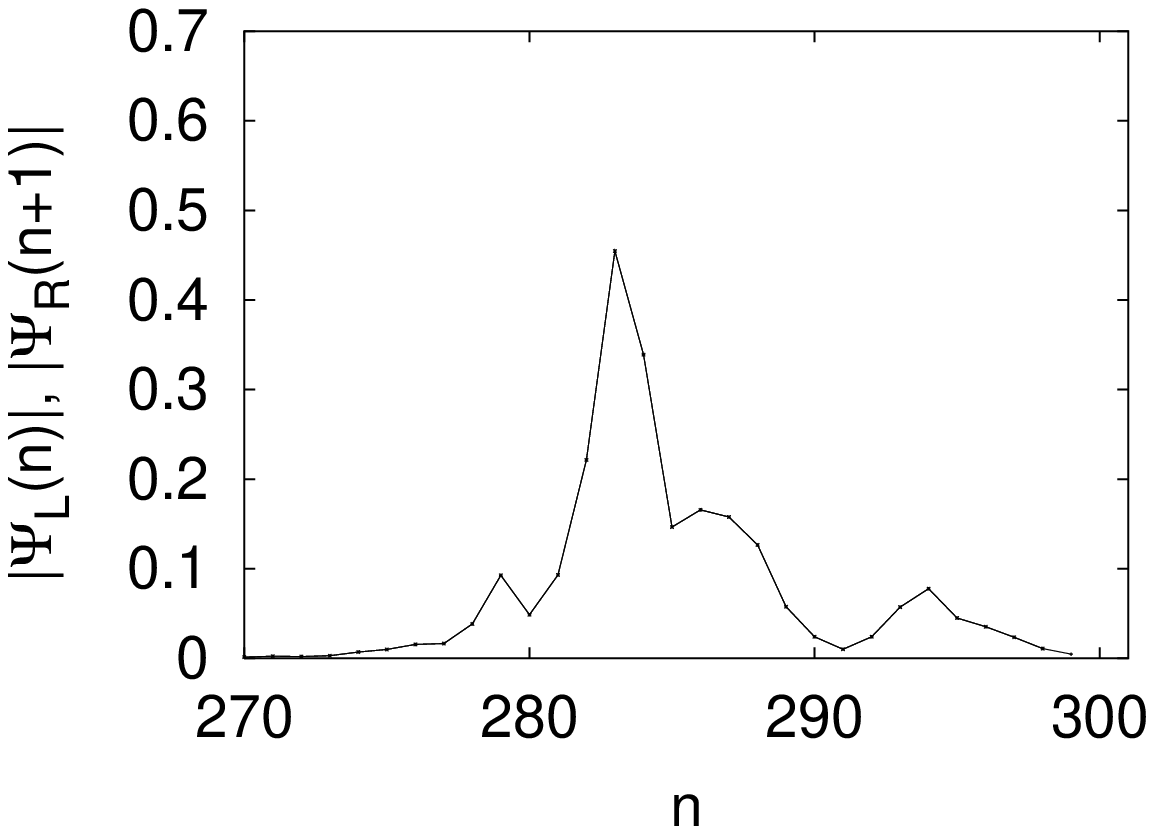} \\
    \includegraphics[scale=0.35]{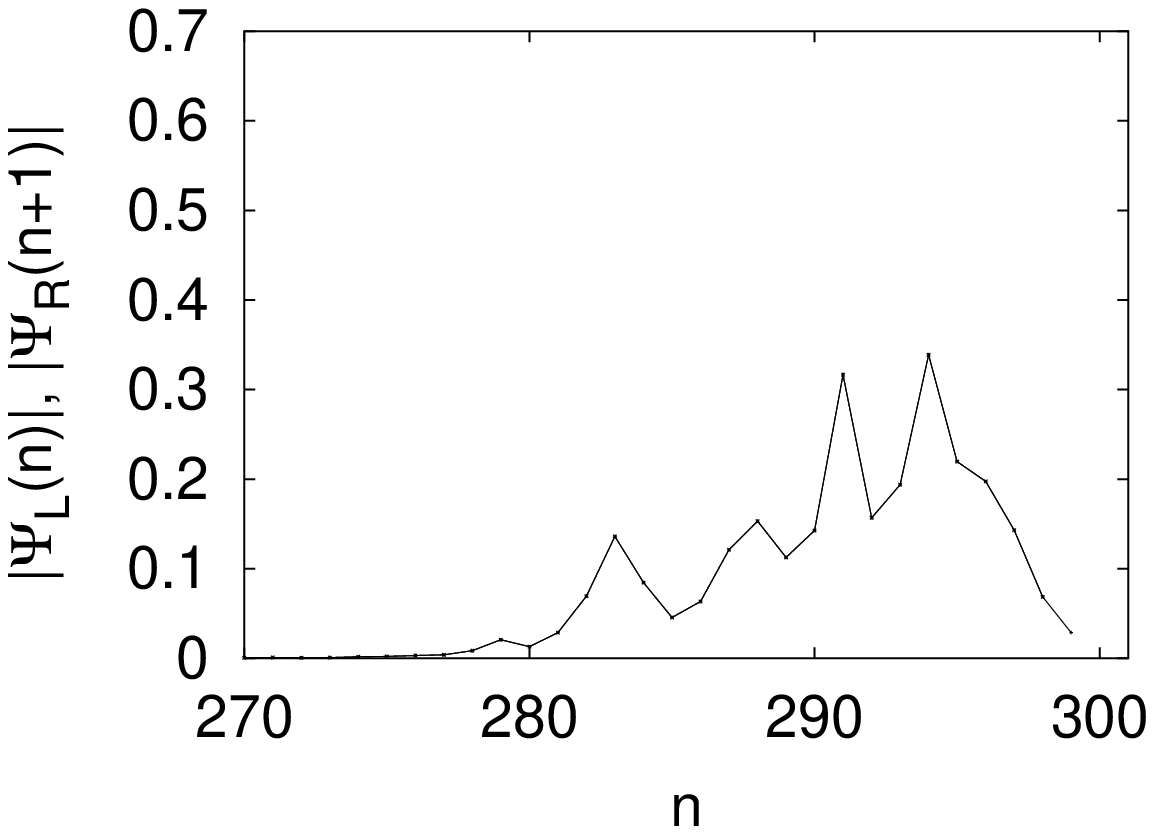} &
    \includegraphics[scale=0.35]{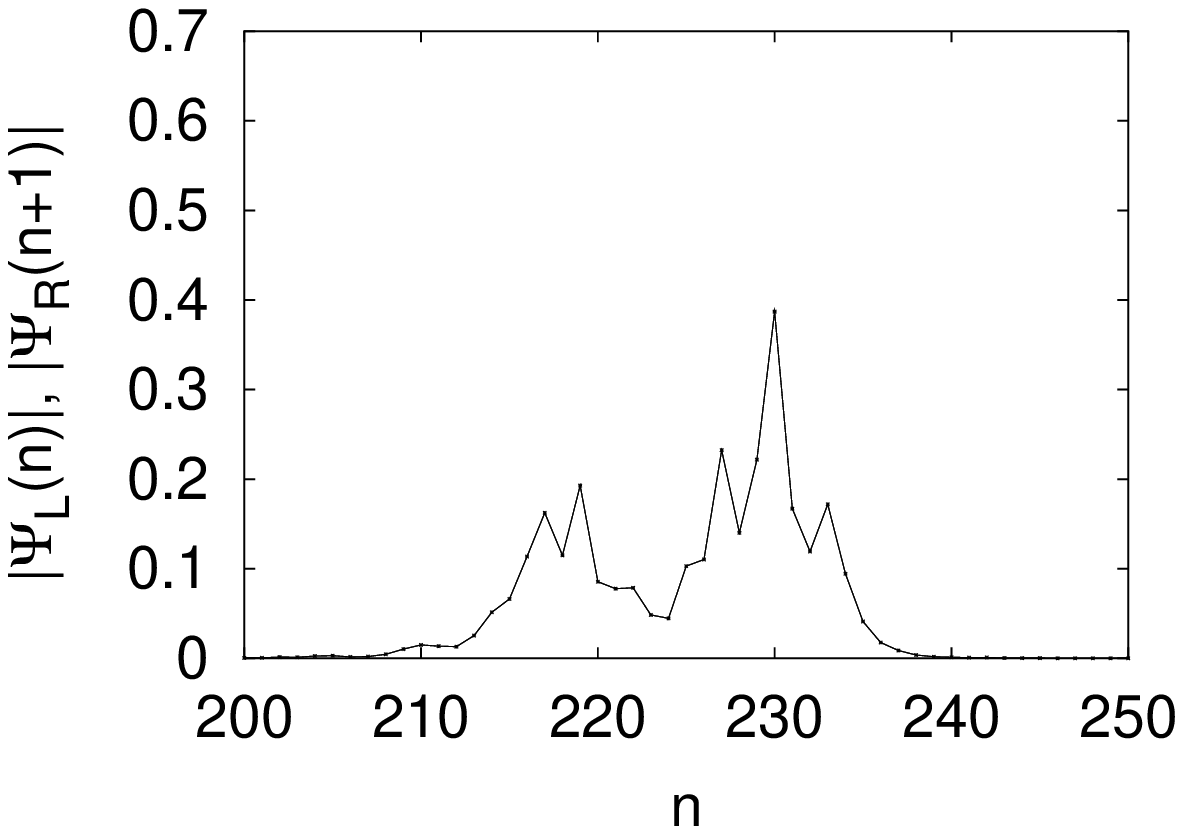} 
  \end{tabular}  
  \caption{The figures show plots of {\em both} $|\Psi_l(n)|$ and
    $|\Psi_r(n+1)|$ superimposed. Fragments of several eigenstates of
    various realisations of quantum random multibakers were chosen
    ($L=101$ in the first figure, $L=301$ in the following figures;
    the parts not shown are zero within the numerical precision).}
  \label{fig:funny}
\end{figure}

Next, we mention an interesting phenomenon.  We first note that the
equations connecting $\widehat{\Psi}_l(n)$ with
$\widehat{\Psi}_l(n-1)$ and $\widehat{\Psi}_l(n+1)$ involve the same
phases and are of the same form as the equation connecting
$\widehat{\Psi}_r(n+1)$ with $\widehat{\Psi}_r(n)$ and
$\widehat{\Psi}_r(n+2)$.  When the transfer matrices are given in the
form of Eq.~(\ref{eq:transf_decomp}) one can show that the two cases
differ only in two random phases, $\phi_1, \phi_2$.  However, since
the equations for $\widehat{\Psi}_l(n)$ and $\widehat{\Psi}_r(n+1)$
separate and have different boundary conditions, there is {\it a
  priori} no connection implied between the solutions to the above
sets of equations. Indeed, after solving the equations numerically for
small $L$ we find that there is no connection between them. Thus it
may come as a surprise that for large~$L$, \(
|\widehat{\Psi}_l(n)|=|\widehat{\Psi}_r(n+1)| \).  This is a numerical
result which accounts for the double peaks in
Figure~\ref{fig:random_states}. To illustrate this equality, we show
several example fragments of eigenstates in Fig.~\ref{fig:funny}.
The equality illustrated here is satisfied to within all the allowed
precision for $L>100$, and is independent of the shape of the
eigenfunction.

The physical explanation is actually quite simple~\footnote{The
  authors are grateful to J. Vollmer for making this nice
  observation.}. The probability current through the boundary between
cells $n$ and $n+1$ is given by $J_{n|n+1} = |\Psi_l(n)|^2 -
|\Psi_r(n+1)|^2$. Therefore the difference between
$|\widehat{\Psi}_l(n)|$ and $|\widehat{\Psi}_r(n+1)|$ implies
non-vanishing current. While such currents can arise in principle one
expects they should be negligible in disordered systems, which leads
to the equality which we observe numerically.

\section{Summary of results and discussion}
\label{sec:summary}

With this paper we have begun our study of transport properties for
quantum multi-baker maps. Our central motivation for this study is, of
course, to use a simple, classically chaotic system with transport and
entropy production and to explore the differences between the
classical system, and its quantum counterpart. The multi-baker map
provides a convenient model for this study.

In order to set the stage for further work, we considered here the
extreme quantum limit of the QMB where each unit cell contains just
two quantum states, in either the position or momentum representation.
Further work will now be devoted to the properties of the QMB for
smaller values of $h$, including a study of the semi-classical regime
where the Planck constant is very small, and the transition to the
classical limit.

In order to explore a variety of versions of the QMB, we used the fact
that the Weyl quantization procedure for maps on torus allows some
freedom in the choice of phases for the quantum states. This freedom
can be thought of as a freedom in the location of the quantum states
in the position and momentum representations, or as the effects of
Bohm-Aharonov current loops on the wave functions in each unit cell.
For large $h$ and for the uniform QMB, we have been able to proceed
with analytical calculations for the quasi-energies, and for the
quantum states for different boundary conditions. These calculations
showed ballistic transport of probability amplitudes within the chain,
due to the Bloch wave structure of the eigenstates, as well as a slow
decay of probability from an open, finite chain, due to the difficulty
that long wavelength modes have in escaping through the exit channels.
The random phase case exhibits localization of wave functions and the
appropriate equations were shown to be close to the basic equations of
the Anderson model of localization. All of these results show that the
behavior of the QMB for $N=2$ is quite different from that of the
classical multi-baker map, which exhibits normal, diffusive transport.
For larger $N$, smaller $h$, we should see behavior that more closely
approximates the classical behavior, at least on the logarithmic time
scale.

There are quite a large number of directions for future studies of the
QMB. For the case of $h=1/2$ we have studied only the uniform and
random phase cases. One can also consider models with periodic
distribution of phases over several cells, models in which there is a
well defined, systematic progression of phases from one cell to the
next, or where the phases in the cells are incommensurate (oscillating
with irrational periods). The situation for larger $N$ is such that
there are $N$ quantum states in each cell, so that there will then be
several transport channels in the QMB. It remains to be seen what the
properties of these systems will be, both as functions of $N$ as well
as functions of the phases of the wave functions in the cells.  As $h$
approaches zero, one can study the semi-classical limit and the
approach to classical, chaotic behavior of the QMB. Clearly, we should
obtain the diffusive behavior independently of the distribution of
phases, that is both for the uniform (ballistic) case, as well as for
the random (localized) case. Such a study should reveal for this
simple system whether the diffusive behavior arises directly
through the semi-classical limit, or whether one needs to assume
additional mechanism --- such as interaction with environment, or
decoherence --- to regain classical properties.  The QMB is simple
enough to consider the clasical limit for uniform phase, random phase,
and incommensurate phase model, among others. The investigation of
these questions will be the subject of further papers in this series.

\begin{acknowledgments}
  This work has been partially supported through the NSF grant
  PHY-98-20824. Conversations with Robert Alicki, Jean Bellissard,
  Mark Fannes, Shmuel Fishman, Pierre Gaspard, Fritz Haake, Salman
  Habib, Ted Kirkpatrick, Shuichi Tasaki, Henk van Beijeren, J\"urgen
  Vollmer, Wojciech \.Zurek, and Karol \.Zyczkowski were of particular
  help. We thank the organizers of the 38th Winter School of
  Theoretical Physics ``Dynamical Semigroups: Dissipation, Chaos,
  Quanta'' in L\c{a}dek Zdr\'oj, Poland, for providing us with an
  opportunity to present and discuss these results.
\end{acknowledgments}

\appendix

\section{Eigenvalues of the open quantum multibaker lie inside the
  unit circle}
\label{sec:app.eigenv.open}

Due to the escape of probability density, the eigenvalues that
determine the time dependence of the probability density in each cell
move to the interior of the unit circle.  A simple proof of this fact
can be given as follows. Let $\Psi(t=0)$ be a normalized eigenstate of
the open quantum multibaker.  Since the multibaker dynamics requires
that \( \widetilde\Psi_b(n-1,t+1) = f_0 \Psi_l(n,t), \,
\widetilde\Psi_t(n+1,t+1) = f_0 \Psi_r(n,t), \) for $n=1,\dots,L-2$,
it follows that
\begin{equation}
  \label{eq:probchange}
  |\widetilde\Psi_b(n-1,t+1)|^2 + |\widetilde\Psi_t(n+1,t+1)|^2 = |\Psi_l(n,t)|^2 +
  |\Psi_r(n,t)|^2. 
\end{equation}
The probability of the system being in cell $n$ is given by \(
\rho(n,t) : = |\Psi_l(n,t)|^2 + |\Psi_r(n,t)|^2 =
|\widetilde\Psi_b(n,t)|^2 + |\widetilde\Psi_t(n,t)|^2,\) where we have
used the unitarity of the transformation $G_2^{-1}$.  In the boundary
cells we have
\begin{equation}
  \left\{
  \begin{array}[c]{rcl}
    |\widetilde\Psi_b(0,t+1)|^2 + |\widetilde\Psi_t(L-1,t+1)|^2 & = & 0,\\
    |\widetilde\Psi_b(1,t+1)|^2 & = & |\Psi_l(0,t)|^2, \\
    |\widetilde\Psi_t(L-2,t+1)|^2 & = & |\Psi_r(L-1,t)|^2.
  \end{array}
  \right.\label{eq:probchangebound}
\end{equation}
By adding up the equations~(\ref{eq:probchange}) for cells
$n=1,2,\dots,L-2$ and~(\ref{eq:probchangebound})  we obtain
\begin{equation}
  \sum_{n=0}^{L-1} \rho(n,t+1) =  \sum_{n=0}^{L-1} \rho(n,t) -
  (|\Psi_r(0,t)|^2 + |\Psi_l(L-1,t)|^2).
\end{equation}
But $\Psi$ is a normalized eigenstate at $t=0$, therefore
\[
  \sum_{n=0}^{L-1} \rho(n,1)
  =|\lambda|^2  
  = 1 -(|\Psi_r(0,0)|^2 + |\Psi_l(L-1,0)|^2).
\]
This implies \( 1-|\lambda|^2 = (|\Psi_r(0,0)|^2 +
|\Psi_l(L-1,0)|^2)\geq 0.\) Thus \( 0\leq |\lambda|^2 \leq 1.  \)
Suppose now $|\lambda|^2 = 1$. Then \( \Psi_r(0,0)=\Psi_l(L-1,0)=0.
\) Thus, from Eq.~(\ref{eq:b9}) and~(\ref{eq:b24}) \( 0 =
\widetilde\Psi_b(0,0) = g_{01} \Psi_l(0,0) \qquad \Rightarrow \qquad
\Psi_l(0,0)=\Psi_r(0,0)=0.  \) It follows that also $\Psi_t(0,0)=0$.
But \( \widetilde\Psi_t(0,0) = \lambda^{-1} \widetilde\Psi_t(0,1) =
\lambda^{-1} f_0 \Psi_r(1,0).  \) Thus $\Psi_r(1,0)=0$. Also, \(
\widetilde\Psi_b(1,1) = f_0\Psi_l(0,0)=0 =\lambda g_{01} \Psi_l(1,0).
\) Therefore $\Psi_l(1,0) = \Psi_r(1,0) = 0$, and so on. Thus the
assumption $|\lambda|^2 = 1$ leads to the eigenstate being identically
0 which cannot be normalized. Therefore all of the eigenvalues lie
inside the unit circle \( 0\leq |\lambda|^2 < 1.\) It is easy to
identify the kernel of $M_L$ since $\lambda=0$ implies \(
|\Psi_r(0,0)|^2 + |\Psi_l(L-1,0)|^2 = 1.  \) Thus the kernel is
spanned by vectors with 0 everywhere apart from $\Psi_r(0)$ and
$\Psi_l(L-1)$.

We are thus led to the conclusion that of the $2L$ eigenstates exactly
two span the kernel, and the eigenvalues corresponding to the
remaining $2L-2$ satisfy $0 < |\lambda| < 1$.  Note that the above
arguments are independent of the distribution of phases and thus apply
also to the open random quantum multibaker.  Generalization to
arbitrary $N$ is obvious.

\section{Anderson model}
\label{app:anderson}

Consider a one-dimensional Schr\"odinger equation on a lattice of
lattice constant $a$ \cite{anderson58s}:
\begin{equation}
  i\hbar \dot{\Psi}_n =
  -\frac{\hbar^2}{2ma^2}[\Psi_{n+1}+\Psi_{n-1}-2\Psi_n] + V_n \Psi_n.
\end{equation}
A time independent equation can be written as \( (\tilde{E} -
\tilde{V}_n) \Psi_n = \Psi_{n+1}+\Psi_{n-1}, \) where $\tilde{E} = 2 -
2ma^2E/\hbar^2$, $\tilde{V}_n = 2ma^2V_n/\hbar^2$. We can also write
it using transfer matrices
\begin{equation}
    \left[
    \begin{array}[c]{c}
      {\Psi}_{n+1}\\ {\Psi}_{n}
    \end{array}    
  \right] = 
  \left[
    \begin{array}[c]{cc}
      \tilde{E} - \tilde{V}_n & -1 \\
      1  & 0 
    \end{array}
  \right] 
  \left[
    \begin{array}[c]{c}
      {\Psi}_{n}\\ {\Psi}_{n-1}
    \end{array}    
  \right],
\end{equation}
which has very similar form to the equation~(\ref{eq:fibonnacci}). A
general discussion of one-dimensional disordered models can be found
in~\cite{crisanti93s,ishii73}

\end{document}